\documentclass[prb,aps,twocolumn,groupedaddress,superscriptaddress]{revtex4-2}
\usepackage[usenames,dvipsnames]{color}
\usepackage[colorlinks=true,linkcolor=Blue,citecolor=Blue,urlcolor=Blue]{hyperref}
\usepackage{graphicx}
\usepackage{bbm}
\usepackage{bm}
\usepackage{amsmath}
\usepackage{amssymb}
\usepackage[version=3]{mhchem}
\usepackage{float}
\usepackage{overpic}
\usepackage{booktabs}
\usepackage{multirow}
\usepackage[normalem]{ulem}

\newcommand{\ba}{\begin{eqnarray}}

\newcommand{\ea}{\end{eqnarray}}
\newcommand{\bd}{\begin{displaymath}}

\newcommand{\nn}{\nonumber \\}

\def\ket#1{|#1\rangle }

\begin{document}

% \title{From Paramagnet to Dipolar Topological Order via Duality and Dipolar SPT}
\title{Gauging Modulated Symmetries via Multiple Gauge Symmetry Operators\\ and Adaptive Quantum Circuits}

\author{Jintae \surname{Kim}}
\email[Electronic address:$~~$]{jintae@illinois.edu}
\affiliation{Department of Physics, Sungkyunkwan University, Suwon 16419, South Korea}
\affiliation{Institute of Basic Science, Sungkyunkwan University, Suwon 16419, South Korea}
\affiliation{Physics Department and Institute of Condensed Matter Theory, University of Illinois at Urbana-Champaign, Urbana, Illinois 61801, USA}

\author{Jong Yeon \surname{Lee}}
\email[Electronic address:$~~$]{jongyeon@illinois.edu}
\affiliation{Physics Department and Institute of Condensed Matter Theory, University of Illinois at Urbana-Champaign, Urbana, Illinois 61801, USA}

\author{Jung Hoon \surname{Han}}
\email[Electronic address:$~~$]{hanjemme@gmail.com}
\affiliation{Department of Physics, Sungkyunkwan University, Suwon 16419, South Korea}

\begin{abstract} 
We introduce an extended framework for the simultaneous gauging of modulated symmetries in $(d+1)$ dimensions, employing {\it multiple} gauge symmetry operators whose corresponding gauging procedures must be carried out simultaneously. Simultaneous gauging can capture a broader class of dualities than sequential gauging, the latter corresponding to the conventional gauging applied in successive steps. In general, performing simultaneous gauging and conventional gauging in sequence constitutes the most general framework for gauging modulated symmetries. We further show that the associated duality transformations can be implemented via adaptive state preparation protocols. As a concrete example, we consider a dipole symmetry in $(2+1)$D and illustrate both the simultaneous gauging procedure and the adaptive preparation protocol. Interestingly, we find that the intermediate state of the simultaneous gauging/adaptive circuit corresponds to a symmetry-protected topological phase protected by the dipole bundle symmetry. Finally, we utilize the duality to analyze the phase diagram of the rank-2 toric code under transverse fields.
\end{abstract}

\date{\today}
\maketitle

\section{Introduction}
(Conventional) gauging is the process of promoting a {\it single} global symmetry generator to a local gauge symmetry, which entails introducing gauge fields as additional dynamical degrees of freedom~\cite{kogut_introduction_1979, mathur_lattice_2016}. Upon gauging a global symmetry generator of a given model and imposing the Gauss's law constraint that enforces local gauge invariance, the theory can be reformulated in terms of dual variables, thereby yielding a dual model. For a global symmetry group generated by multiple symmetry generators, one can perform the gauging procedures in a sequential manner, which we refer to as sequential gauging.

Recently, the notion of global symmetry has been significantly generalized to encompass a broader class, including higher-form symmetries~\cite{mcgreevy_generalized_2023, gomes_introduction_2023, brennan_introduction_2023}, subsystem symmetries~\cite{you_higher-order_2024,you_higher-order_2018}, modulated symmetries~\cite{sala22,han24,you24, bulmash_defect_2025,kim25}, and non-invertible symmetries~\cite{shao_whats_2023, brennan_introduction_2023, bhardwaj24}, thereby providing a unified framework for classifying and investigating unconventional phases of matter~\cite{nandkishore-fracton-review, yizhiyou20,gromov24, seifnashri_cluster_2024, kim_model_2021, kim_hybrid_2022, lee_decoding_2022, xu_entanglement_2025, lam24, choi_non-invertible_nodate, gorantla_tensor_2024, ebisu_foliated_2024, huang23, seiberg_exotic_2020,seiberg_exotic_2021}. 
Moreover, the approach of understanding various phases through the gauging of generalized symmetries has garnered substantial attention~\cite{bulmash_gauging_2019,prem_gauging_2019,cheng_gauging_2022, shirley_foliated_2019, delfino23b, kuno_interplay_2023,ebisu24,gorantla_string-membrane-nets_2025,cao24, pace_gauging_2024, ebisu25,li_non-invertible_2024,seifnashri_gauging_2025,lyons_protocols_2024, kim25, li_symmetry-enriched_2023,yoshitome_generalized_2025}.

In particular, a method for gauging modulated symmetries using a {\it single} gauge symmetry operator has been recently proposed~\cite{cao24,pace_gauging_2024, ebisu25}. For example, one can gauge the dipole symmetry and the global charge symmetry simultaneously, a procedure we refer to as {\it simultaneous gauging}. This suggests that simultaneous gauging provides a new framework for implementing symmetry gauging, enabled by the intertwined nature of these symmetries under translation. In this Letter, we extend this idea to the use of {\it multiple} gauge symmetry operators for the simultaneous gauging of more general modulated symmetries. Note that our new scheme is subject to constraints on the gauge symmetry operators that ensure the simultaneity of the gauging procedure. We refer to the case involving $n$ gauge symmetry operators as {\it $n$-simultaneous gauging}, with the case $n\,{=}\,1$ studied previously~\cite{cao24,pace_gauging_2024, ebisu25}. In general, performing simultaneous gauging and conventional gauging in sequence forms the most general framework for gauging modulated symmetries. To illustrate the proposed construction, we present $1$-, $2$-, and $n$-simultaneous gauging examples in (2+1) dimensions associated with dipole symmetries~\cite{ma18, bulmash18,oh22a,pace-wen, oh22b, oh23, kim23, watanabe23, ebisu23a, delfino23b, gorantla23, delfino23, huang23,ebisu24,han24, sachdev02,morningstar20,sala20,feldmeier20,gromov20,surowka22,gorantla22,sala22,seiberg23,skinner23,dSPT,lake22,zechmann23,lake23,ro24}, which are of increasing importance given the feasibility of experimental realizations in tilted optical lattices~\cite{bakr20,aidelsburger21,weitenberg22}. The dual models associated with the 1- and 2-simultaneous gauging are the anisotropic dipolar toric code~\cite{ebisu23a} and the rank-2 toric code (R2TC)~\cite{oh22a, pace-wen, oh22b, oh23, kim23}, respectively.

A closely related question is whether the duality transformation associated with $n$-simultaneous gauging can be implemented by the adaptive quantum circuits. Previous studies have demonstrated that protocols of adaptive state preparation are closely linked to dualities between paramagnetic and topological phases~\cite{nathanan20,verresen21a}. In particular, for topologically ordered states such as the toric code, various measurement-based preparation schemes have been proposed~\cite{Hastings21,cirac21,verresen21a,verresen21b,bravyi22,hsieh22,aasen22,nathanan23, raussendorf01,raussendorf05}. Furthermore, significant progress has recently been made in the implementation of adaptive quantum circuits~\cite{Pino21,stutz21,gambetto21,isaackim23,iqbal24,ren_efficient_2025}, suggesting a viable path toward the dynamic realization of highly entangled quantum states.

To address this question, we present a concrete procedure for the adaptive preparation of the R2TC. Moreover, we point out that the general $n$-simultaneous gauging can be implemented using an adaptive quantum circuit without incurring additional complexity compared to sequential gauging. 
We show that the intermediate state of the simultaneous gauging/adaptive circuit of the R2TC exhibits an interesting symmetry-protected topological (SPT) phase protected by dipolar bundle symmetry. Furthermore, using the duality enabled by the simultaneous gauging, we analyze the phase transition associated with the R2TC, demonstrating the utility of the proposed framework.

\section{Modulated symmetries and simultaneous gauging}

We define the $\mathbb{Z}_N$ Pauli operators acting on the qudit at position $\bf r$ as $Z_{\bf r} |s_{\bf r}\rangle_{\bf r} = \omega^{s_{\bf r}} |s_{\bf r}\rangle_{\bf r}$ and $X_{\bf r} |s_{\bf r}\rangle_{\bf r} = |s_{\bf r} + 1\rangle_{\bf r}$, where $\omega = \exp(2\pi i / N)$. Translational symmetry is assumed in each of the $d$ spatial directions, with unit lattice vectors $\hat{a}$ $(a=1,2,\ldots, d)$, and the length of each dimension denoted by $L_a$. We consider multiple qudits within a unit cell in $(d+1)$D, labeled by $j$. The position of qudit $j$ is given by ${\bf r}_j = {\bf o}_j + \sum_{a} a_j \hat{a}$, where $a_j$ is an integer and ${\bf o}_j$ denotes its relative position within the unit cell.

The overall symmetry group $G$ can be written as
\begin{align}
G = G_{\rm mod} \rtimes G_{T},
\end{align}
where $G_{\rm mod}$ denotes the Abelian group of modulated symmetries, and $G_T$ the group of translation symmetries~\cite{pace_gauging_2024, bulmash_defect_2025}. The generators $g_\alpha$ of the modulated symmetry group $G_{\rm mod}$ are assumed to be $\mathbb{Z}_N$ operators for simplicity.

We define $n$-simultaneous gauging as the procedure of gauging multiple symmetries using $n$ gauge symmetry operators whose corresponding gauging procedures must be carried out simultaneously. To illustrate $n$-simultaneous gauging of modulated symmetries with explicit examples, we make the choice of defining a gauge symmetry operator $h_{{\bf r}_j}$ for each site ${\bf r}_j$, which acts $X$ on the site ${\bf r}_j$ and other operations on the gauge degrees of freedom~\footnote{The choice of $X$ in $h_{{\bf r}_j}$ is natural since we restrict to $\mathbb{Z}_N$ symmetry generators. In general, the natural choice can be $X$ raised to some power.}. This means there are $n$ qudits within a unit cell and $n$ gauge symmetry operators. The relations between $g_{\alpha}$ and $h_{{\bf r}_j}$ are then given by
\begin{align}
g_\alpha \equiv  \prod_{j=1}^n\prod_{{\bf r}_j} X_{{\bf r}_j}^{f_{{\bf r}_j}^{(\alpha)}}= \prod_{j=1}^n  \prod_{{\bf r}_j} h_{{\bf r}_j}^{f_{{\bf r}_j}^{(\alpha)}},
\label{eq:symg}
\end{align}
where $f_{{\bf r}_j}^{(\alpha)} \in \mathbb{Z}_N$ are assumed to be position-dependent functions and satisfying $f_{{\bf r}_j+L_a\hat{a}}^{(\alpha)}=f_{{\bf r}_j}^{(\alpha)}$ for $a=1,2,\ldots,d$. This corresponds to dipole symmetry when $f_{{\bf r}_j}^{(\alpha)}$ is a linear function of ${\bf r}_j$. Simultaneity is ensured when any nonempty proper subset $O$ of gauge symmetry operators $h_{{\bf r}_j}$ share gauge fields with its complement $O^c$. In the case of two gauge symmetry operators, this condition simply requires that they share gauge fields. 

In the following examples, we focus on modulated symmetry groups that satisfy the constraints of gauge symmetry operators, as well as on $(2+1)$D cases that clearly demonstrate that the duality resulting from simultaneous gauging can realize models that cannot be obtained through sequential gauging~\cite{pace_gauging_2024} (see Appendix~\ref{appendix:A}). Note that the overall duality obtained through sequential gauging can be realized directly through simultaneous gauging.
\\

\subsection{1-simultaneous gauging}

In this case, a number of symmetry generators $g_\alpha$ are generated by a single gauge symmetry operator~\footnote{Typically, the gauging of distinct (non-modulated) symmetry generators is performed using a set of independent gauge symmetry operators, which are applied sequentially.}. 
A simple illustrative example is given on a square lattice with $m\,{=}\,1$ and ${\bf o}_1$ taken to be the zero vector. We consider both the charge and the dipole ($x$-direction) $\mathbb{Z}_N$ symmetries, together with the corresponding gauge symmetry operators given as
\begin{align}
g_1 & \equiv \prod_{{\bf r}} X_{{\bf r}} = \prod_{{\bf r}} h_{{\bf r}}, ~~~~~~~g_2  \equiv \prod_{{\bf r}} X_{\bf r}^{x} = \prod_{{\bf r}} h_{{\bf r}}^{x}
\end{align}
where ${\bf r}=x \hat{x}+y\hat{y}$. $h_{{\bf r}}$ can be chosen as
\begin{align}
h_{{\bf r}} & \equiv X_{{\bf r}} \bar{Z}_{{\bf r}+\hat{x}}^{-1} \bar{Z}^{2}_{{\bf r}} \bar{Z}^{-1}_{{\bf r} -\hat{x}} \bar{Z}^{-1}_{{\bf r} +\hat{y}/2} \bar{Z}_{{\bf r} -\hat{y}/2},
\end{align}
where the gauge fields $\bar{Z}$ live on the vertices and vertical edges. The corresponding duality can be summarized as
\begin{align}
&X_{{\bf r}} 
\rightarrow \bar{Z}_{{\bf r}+\hat{x}} \bar{Z}^{-2}_{{\bf r}} \bar{Z}_{{\bf r}-\hat{x}} \bar{Z}_{{\bf r}+\hat{y}/2} \bar{Z}^{-1}_{{\bf r}-\hat{y}/2}\nn
&Z_{{\bf r}+\hat{x}} Z^{-2}_{{\bf r}} Z_{{\bf r}-\hat{x}}  \rightarrow \bar{X}^\dagger_{\bf r}\nn
&Z_{{\bf r}}^{-1} Z_{{\bf r}+\hat{y}} \rightarrow \bar{X}_{{\bf r}+\hat{y}/2}.\label{eq:ar2tcG}
\end{align}
Under this duality, the paramagnet Hamiltonian $H = \sum_{\bf r} X_{\bf r} + {\rm h.c.}$, together with the zero-flux condition, maps to an anisotropic dipolar toric code~\cite{ebisu23a}, where the $e$ and $m$ anyons are showing dipolar character along $x$-direction.

\begin{figure}[ht]
\centering
\includegraphics[width=0.9\columnwidth]{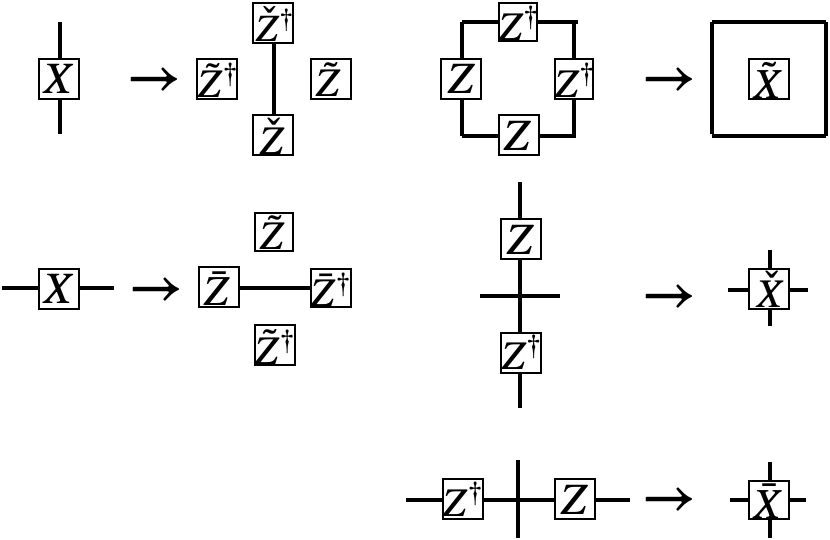}
\caption{The duality transformation of operators for the $2$-simultaneous gauging of $g_{\rm d}$, $g_{e_{\rm h}}$, and $g_{e_{\rm v}}$ is depicted. A gauge field $\tilde{Z}$ is defined at the centers of the plaquettes. Two additional gauge fields, $\bar{Z}$ and $\check{Z}$, are assigned to the lattice vertices. } 
\label{fig:R2TCd}
\end{figure}

\subsection{2-simultaneous gauging}

Consider a square lattice with two qudits per unit cell ($m\,{=}\,2$) with ${\bf o}_1=\frac{1}{2} \hat{x}$, ${\bf o}_2=\frac{1}{2}\hat{y}$, i.e., qudits reside on the edges. Two $\mathbb{Z}_N$ charge symmetry generators, $g_{e_{\rm h}}$ and $g_{e_{\rm v}}$, together with a $\mathbb{Z}_N$ dipole symmetry generator $g_{\rm d}$, can be represented in terms of the corresponding gauge symmetry operators as
\begin{align}
g_{e_{\rm h}}&\equiv\prod_{e_{\rm h}} X_{e_{\rm h}} = \prod_{e_{\rm h}} h_{e_{\rm h}}, ~~
g_{e_{\rm v}}\equiv\prod_{e_{\rm v}} X_{e_{\rm v}} = \prod_{e_{\rm v}} h_{e_{\rm v}} ,\nn
g_{\rm d} &\equiv \prod_{e_{\rm h} } X_{e_{\rm h}}^{-y_{e_{\rm h}}} \prod_{e_{\rm v}} X_{e_{\rm v}}^{x_{e_{\rm v}}} = \prod_{e_{\rm h}} h_{e_{\rm h}}^{-y_{e_{\rm h}}} \prod_{e_{\rm v}} h_{e_{\rm v}}^{x_{e_{\rm v}}} ,
 \label{eq:symd}
\end{align}
where $e_{\rm h}$ and $e_{\rm v}$ denote qudits residing on horizontal and vertical edges, respectively. Here, $g_{\rm d}$ is given by the product of the dipole symmetry along the $(-y)$-direction for horizontal qudits and the dipole symmetry along the $x$-direction for vertical qudits.

We emphasize that the $h_{e_{\rm h}}$ and $h_{e_{\rm v}}$ must share gauge fields, as their combined action is required to reproduce only $g_{e_{\rm h}}$, $g_{e_{\rm v}}$, and $g_{\rm d}$. Otherwise, the product of $h_{e_{\rm h}}$ and $h_{e_{\rm v}}$ may also generate operators such as $\prod_{e_{\rm h}} X_{e_{\rm h}}^{-y_{e_{\rm h}}}$ and $\prod_{e_{\rm v}} X_{e_{\rm v}}^{x_{e_{\rm v}}}$, which are not the symmetries under consideration. Therefore, the gauging of the three symmetries must be performed simultaneously.

The gauge symmetry operators $h_{e_{\rm h}}$ and $h_{e_{\rm v}}$ can be chosen as
\begin{align}
h_{e_{\rm h}}&= X_{e_{\rm h}} \bar{Z}_{e_{\rm h}-\frac{1}{2}\hat{x}} \bar{Z}_{e_{\rm h}+\frac{1}{2}\hat{x}}^\dag \tilde{Z}_{e_{\rm h}-\frac{1}{2}\hat{y}}^\dag \tilde{Z}_{e_{\rm h}+\frac{1}{2}\hat{y}} \nn
h_{e_{\rm v}}&=X_{e_{\rm v}} \check{Z}_{e_{\rm v}-\frac{1}{2}\hat{y}} \check{Z}_{e_{\rm v}+\frac{1}{2}\hat{y}}^\dagger \tilde{Z}_{e_{\rm v}-\frac{1}{2}\hat{x}}^\dag \tilde{Z}_{e_{\rm v}+\frac{1}{2}\hat{x}} .
\label{hev-and-heh}
\end{align}
The gauge fields defined at the plaquette centers are denoted by $\tilde{Z}$. Two additional gauge fields, $\bar{Z}$ and $\check{Z}$, are defined at the vertices—see Fig.~\ref{fig:R2TCd}. The same plaquette-centered gauge fields $\tilde{Z}$ appearing in both $h_{e_{\rm h}}$ and $h_{e_{\rm v}}$. The duality transformation of operators is depicted in Fig.~\ref{fig:R2TCd}.

Applying the duality transformation in Fig.~\ref{fig:R2TCd}, the paramagnet Hamiltonian, together with the zero-flux condition, maps to the R2TC~\cite{oh22a,pace-wen,oh22b,oh23, kim23} whose anyonic excitations exhibit fractonic behavior. 

\begin{figure*}
\setlength{\abovecaptionskip}{5pt}
\centering
\begin{overpic}[width=2.02\columnwidth]{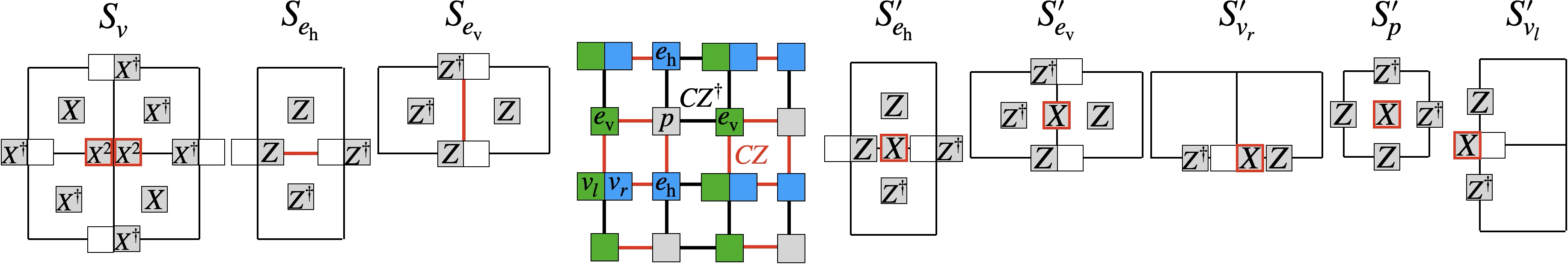} 
\put(-1,16){(a)}
\put(35,16){(b)}
\put(52,16){(c)}
\end{overpic}
\caption{(a) Three stabilizers of the R2TC. Red squares and lines indicate the stabilizer location.
(b) Qudits and the quantum circuit arrangements for the preparation of dCS. Two qudits are placed at the vertices and labeled $v_{l}$ and $v_r$, respectively. Single qudits are placed at the horizontal and vertical edges ($e_{\rm h}$, $e_{\rm v}$) as well as plaquette centers ($p$). The entanglers $(CZ)_{ab}$ and $(CZ)_{ab}^{\dagger}$ act on the two qudits connected by red and black lines, respectively, in accordance with the rule that the $\textrm{CZ}$ operation involving one of the vertex qudits takes place between qudits of the same color. (c) Five stabilizers that realize the dCS. Red squares indicate the stabilizer location.}
\label{fig:stab}
\end{figure*}

\subsection{$n$-simultaneous gauging}

Now we consider a square lattice with $n$ qudits per unit cell, where ${\bf o}_j = \tfrac{j-1}{n}\hat{x}$ ($1 \le j \le n$). We introduce $\mathbb{Z}_N$ charge symmetry generators for each $j$ and a single $\mathbb{Z}_N$ dipole ($x$-direction) symmetry generator acting on all qudits, which are related to the gauge symmetry operators through
\begin{align}
g_{j}&\equiv\prod_{{\bf r}_j} X_{{\bf r}_j}=\prod_{{\bf r}_j} h_{{\bf r}_j} \quad \textrm{for $j=1,...,n$}\nn
 g_{\rm d}&\equiv\prod_{j=1}^{n} \Big[ \prod_{{\bf r}_j} X_{{\bf r}_j}^{x_j} \Big] = \prod_{j=1}^{n} \Big[ \prod_{{\bf r}_j} h_{{\bf r}_j}^{x_j}\Big].\label{eq:symn}
\end{align}
As in the case of $2$-simultaneous gauging, if there exists a nonempty proper subset $O$ of gauge symmetry operators that does not share gauge fields with its complement $O^c$, then the product of the gauge symmetry operators in $O$ generates an operator that is not a symmetry generator. Consequently, such subset $O$ does not exist and these symmetries can only be gauged simultaneously.

Accordingly, we can make the following choice:
\begin{widetext}
\begin{align}
h_{{\bf r}_1}=&X_{{\bf r}_1} \left( \bar{Z}^{(1)}_{{\bf r}_1-\hat{x}}  [\bar{Z}^{(1)}_{{\bf r}_1}]^{-2} \bar{Z}^{(1)}_{{\bf r}_1+\hat{x}}\right) \left( \bar{Z}^{(2)}_{{\bf r}_1+(\frac{1}{2n}-1)\hat{x}} [\bar{Z}^{(2)}_{{\bf r}_1+\frac{1}{2n}\hat{x}}]^{-1}\right) \left( [\bar{Z}^{(1)}_{{\bf r}_1-\frac{1}{2}\hat{y}}]^{-1} \bar{Z}^{(1)}_{{\bf r}_1+\frac{1}{2}\hat{y}} \right) \nn
h_{{\bf r}_j}=&X_{{\bf r}_j} \left( [\bar{Z}^{(2j-2)}_{{\bf r}_j-\frac{1}{2n}\hat{x}}]^{-1} \bar{Z}^{(2j-2)}_{{\bf r}_j+(1-\frac{1}{2n})\hat{x}} \right) \left( \bar{Z}^{(2j-1)}_{{\bf r}_j-\hat{x}} [\bar{Z}^{(2j-1)}_{{\bf r}_j}]^{-2} \bar{Z}^{(2j-1)}_{{\bf r}_j+\hat{x}} \right) \left(\bar{Z}^{(2j)}_{{\bf r}_j+(\frac{1}{2n}-1)\hat{x}} [\bar{Z}^{(2j)}_{{\bf r}_j+\frac{1}{2n}\hat{x}}]^{-1} \right) \left( [\bar{Z}^{(2j-1)}_{{\bf r}_{j}-\frac{1}{2}\hat{y}}]^{-1} \bar{Z}^{(2j-1)}_{{\bf r}_{j}+\frac{1}{2}\hat{y}}\right)\nn
h_{{\bf r}_n}=&X_{{\bf r}_n}\left( [\bar{Z}^{(2n-2)}_{{\bf r}_n-\frac{1}{2n}\hat{x}}]^{-1} \bar{Z}^{(2n-2)}_{{\bf r}_n+(1-\frac{1}{2n})\hat{x}} \right)\left([\bar{Z}^{(2n-2)}_{{\bf r}_n-\frac{1}{2}\hat{y}}]^{-1} \bar{Z}^{(2n-2)}_{{\bf r}_n+\frac{1}{2}\hat{y}} \right),
\end{align}
\end{widetext}
where $2\leq j \leq n-1$, and a total of $2n-2$ distinct gauge fields $\bar{Z}^{(1)}$ through $\bar{Z}^{(2n-2)}$ are introduced. As expected, each $\bar{Z}^{(j)}$ appears in multiple gauge symmetry operators $h^{(j)}$ in such a way that the gauge symmetry operator constraint is satisfied. The several examples we discussed so far suggest that $n$-simultaneous gauging applies to a broad class of modulated symmetries and can be extended to general $(d+1)$D.

\section{Adaptive preparation and $n$-simultaneous gauging}

We now illustrate how the duality associated with the $2$-simultaneous gauging in $(2+1)$D can be formulated in the language of adaptive circuits, then make comments on generalizations to arbitrary $n$ and $(d+1)$D.

The dual model obtained from the 2-simultaneous gauging is the R2TC, which is a stabilizer model on a square lattice with three stabilizers $S_v$, $S_{e_{\rm h}}$, and $S_{e_{\rm v}}$ shown in Fig.~\ref{fig:stab}(a). Two qudits are placed on each vertex, and one qudit is at the plaquette center~\footnote{These stabilizers are equivalent to those in \cite{oh22a, kim23}, up to qudit rotations, and are more elegant considering the duality interpretation.}. Further details of the model can be found in \cite{oh23} and in Appendix~\ref{appendix:B}.

To prepare the ground state of R2TC in the adaptive scheme, we first apply some unitary gates to construct the intermediate state, hereby denoted the dipolar cluster state (dCS). The dCS is defined on a square lattice with five qudits in a unit cell as in Fig.~\ref{fig:stab}(b). To prepare the dCS, each qudit is initialized in the state $|\overline{0}\rangle$, $X|\overline{0}\rangle = |\overline{0}\rangle$. A pairwise $\textrm{CZ}$ (CZ$^\dagger$) operation, defined by $(CZ)_{ab} |s_a , s_b\rangle = \omega^{s_a s_b}|s_a , s_b\rangle$, $s_{a,b} \in \mathbb{Z}_N$, takes place between two qudits connected by red (black) lines in Fig.~\ref{fig:stab}(b). The $\textrm{CZ}$ operation involving one of the vertex qudits takes place between qudits of the same color. The paramagnetic state is stabilized by $X$ acting on each qudit. Therefore, the stabilizers that realize dCS are obtained by performing the above-mentioned CZ operations, and invoking the identity $(CZ)_{ab} X_a (CZ)_{ab}^\dagger = X_a Z_b$. The five stabilizers $S'_{e_{\rm h}}$, $S'_{e_{\rm v}}$, $S'_{v_r}$, $S'_{p}$, and $S'_{v_l}$ thus obtained are shown in Fig.~\ref{fig:stab}(c). 

To realize the ground state of R2TC from the dCS, we project all the qudits located at the edges to the $X = +1$ eigenstate. The remaining qudits at the vertices and plaquette centers then form the ground state of R2TC. The relation to gauging arises from the fact that the two stabilizer of dCS, $S'_{e_{\rm h}}$ and $S'_{e_{\rm v}}$, are precisely the gauge symmetry operators $h_{e_{\rm h}}$ and $h_{e_{\rm v}}$ for the 2-simultaneous gauging of $g_{\rm d}$, $g_{e_{\rm h}}$, and $g_{e_{\rm v}}$ mentioned earlier. The other three stabilizers of dCS, $S'_{v_r}$, $S'_{p}$, and $S'_{v_l}$, do not participate in the gauging {\it per se}, but a carefully chosen product of them realizes the flux-free condition enforcing $S_v = 1$ in the resulting adaptive state. Measurements (projection) on the edge qudits amount to performing the duality, eliminating the matter fields and keeping only the gauge fields on the un-measured sites.

The R2TC example outlines the general principle for realizing $n$-simultaneous gauging in $(d\,{+}\,1)$D through adaptive circuits. First, we prepare `physical' qudits on matter field sites and ancilla qudits on gauge field sites, stabilized respectively by $X$ and $\bar{X}$. Let us call the overall initial state $|+\rangle$. Through suitably chosen unitary rotations $U$, which can always be constructed, each $X_{{\bf r}_j}$ (${\bf r}_j \in$ vertices) becomes $h_{{\bf r}_j}$ in the given gauging scheme, i.e. $U X_{{\bf r}_j} U^\dag = h_{{\bf r}_j}$, and the overall state $U|+\rangle$ is stabilized by various $h_{{\bf r}_j}$. At the same time, $\bar{X}_{{\bf r}_a}$ at the ancilla sites ${\bf r}_a$ undergoes its own transformation $U \bar{X}_{{\bf r}_a} U^\dag = \bar{h}_{{\bf r}_a}$ and also stabilizes the intermediate state $U|+\rangle$. 

The next stage is the projection (measurement) on the vertex qudits to $X_{{\bf r}_j} =+1$. This process commutes with $h_{{\bf r}_j}$, but not with $\bar{h}_{{\bf r}_a}$ in general. Nonetheless, an appropriately chosen product of $\bar{h}_{{\bf r}_a}$ operators can commute with the projection and impose the flux-free condition. In the process, each symmetry generator $g_\alpha$ becomes a product of operators in the ancilla (dual) space which encodes conservation laws for quasiparticles in the dualized model (See Appendix~\ref{appendix:C}). Ultimately, we expect that $n$-simultaneous gauging can be implemented as an adaptive quantum circuit in arbitrary $(d+1)$D systems with no extra difficulty compared to the realization of sequential gauging.
\\

\section{Dipolar cluster state}

The intermediate state arises during the process of an adaptive quantum circuit. It turns out that this state possesses notable characteristics worthy of independent discussion. This point is exemplified through the analysis of the dCS.

First of all, the dCS can be viewed as an SPT state protected by (i) two charge $\mathbb{Z}_N$ symmetries $g_{e_{\rm h}}$ and $g_{e_{\rm v}}$, (ii) a dipole bundle symmetry $g_{\rm d}$, and (iii) three 1-form symmetries, $g_X^1$, $g_X^2$, and $g_X^3$, which are also the three 1-form symmetries of R2TC. The notion of dipole bundle symmetry had been introduced to account for the subtlety in defining a dipole symmetry operator in a closed chain, and can generalize to arbitrary lattice with periodic boundary conditions~\cite{dSPT}. The SPT state protected by dipole symmetry is referred to as dipolar SPT (dSPT). The dCS state is, to our knowledge, the first explicit construction of dSPT state in two dimensions.

Secondly, the dSPT hosts symmetry-protected boundary states arising from mutual anomalies among its governing symmetries. Using the bulk stabilizers of dCS, the action of global symmetries on the dCS ground state $|\Psi\rangle$ can be shown to localize to the upper (u) and lower (l) boundaries as $g_{\rm a} |\Psi \rangle = g_{\rm a}^{\rm u} g_{\rm a}^{\rm l}|\Psi \rangle~ ({\rm a= h, v, d})$, where
\begin{align}
g_{\rm h}^{\rm u} &= \prod_{p,e_{\rm h} \in {\rm u}} Z_p^\dagger X_{e_{\rm h}},~~
g_{\rm v}^{\rm u} =\prod_{v_l \in {\rm {\rm u}}} Z_{v_{l}},~~ g_{\rm d}^{\rm u}=\prod_{v_l \in {\rm u}} Z_{v_l}^{x_{v_{l}}},
\label{upper-boundary-symmetry}
\end{align}
represent fractionalized symmetry operators on the upper boundary. (Similar operators exist for the lower boundary.) Assuming a semi-infinite system with only the upper boundary, the three operators derived in \eqref{upper-boundary-symmetry} serve as the symmetry operators of the edge-localized modes. They are independent symmetries, in the sense that no two operators can be related by the product of stabilizers in the bulk. Furthermore, these fractionalized operators have nontrivial commutation relations with the 1-form symmetries defined along the $y-$axis: 
\begin{align}
g_{\rm h}^{\rm u} g_X^2 &= \omega g_X^2 g_{\rm h}^{\rm u},~~ g_{\rm v}^{\rm u} g_X^1 = \omega g_X^1 g_{\rm v}^{\rm u},~~ g_{\rm d}^{\rm u} g_X^3 = \omega g_X^3 g_{\rm d}^{\rm u} ,
\label{anomaly-relations}
\end{align}
implying that the action by $g_X^2 , g_X^1, g_X^3$ on a ground state toggles the eigenvalues of $g_{\rm h}^{\rm u}, g_{\rm v}^{\rm u} , g_{\rm d}^{\rm u}$ by $\omega$. Therefore, we have $N^3$ boundary zero modes labeled by $(q_{\rm h}, q_{\rm v}, q_{\rm d}) \in \mathbb{Z}_N^3$~\footnote{Although the commutation relation between $g_d^{\rm u}$ and $g_X^1$ is nontrivial, it does not affect the conclusion because of the existence of $g_X^3$.}. For a periodic lattice in the $x$ direction of length $L_x$, the proper symmetry operator is $g_{\rm d}^{N/\gcd(L_x, N)}$ instead of $g_{\rm d}$, resulting in $N^2\gcd(L_x, N)$-fold degeneracy. 

The case for more general boundary types and a complementary analysis based on explicit construction of symmetric boundary Hamiltonian is presented in Appendix~\ref{appendix:E}. As already exemplified by $1$- and $2$-simultaneous gauging, we anticipate the appearance of a variety of SPT phases protected by modulated symmetries upon following the $n$-simultaneous gauging procedure.

\emph{Quantum phase transitions}.--- By extending the understanding of a broader class of topological phases through the $n$-simultaneous gauging, the analysis of phase transitions becomes more tractable. We will utilize the duality in Fig.~\ref{fig:R2TCd} to understand the phase transition of R2TC subject to various transverse fields: 
\begin{align} 
&H(\lambda_1, \lambda_2, \lambda_3)=-\sum_{v} S_{v}-\sum_{e_{\rm h}} S_{e_{\rm h}}-\sum_{e_{\rm v}} S_{e_{\rm v}}\nn
&~-\lambda_1\sum_{p} X_p - \lambda_2 \sum_v X_{v_l} -\lambda_3 \sum_v X_{v_r}  +{\rm h.c.}.
\label{eq:H123} 
\end{align}
The transverse field terms $X_{v_l}$, $X_{v_r}$, and $X_{p}$ generate quasiparticles (anyons) associated with $S_{e_{\rm v}}$, $S_{e_{\rm h}}$, and both $S_{e_{\rm v}}$ and $S_{e_{\rm h}}$, respectively. The anyons associated with $S_{e_{\rm h}}$ and $S_{v}$ exhibit nontrivial mutual statistics, as do those associated with $S_{e_{\rm v}}$ and $S_{v}$~\cite{oh22a}. 

The model $H(\lambda_1, \lambda_2, \lambda_3)$ hosts four distinct phases $P_1$ through $P_4$. The details of $P_1$ to $P_4$ phases can be found in Appendix~\ref{appendix:F}.

We can dualize the Hamiltonian \eqref{eq:H123} by performing the inverse of the map given in Fig.~\ref{fig:R2TCd}:
\begin{align}
& H_{\rm dual}(\lambda_1,\lambda_2,\lambda_3)=-\sum_{e_{\rm h}} X_{e_{\rm h}}-\sum_{e_{\rm v}} X_{e_{\rm v}}\nn
&~-\lambda_1\sum_{p} S_p'' -\lambda_2\sum_{v_l} S_{v_l}'' -\lambda_3 \sum_{v_r} S_{v_r}'' +{\rm h.c.},\label{eq:dualH}
\end{align}
where $S_p''$, $S_{v_l}''$, and $S_{v_r}''$ correspond to $S_p'$, $S_{v_l}'$, and $S_{v_r}'$ with the $X$ operator omitted, respectively (they also appear on the left side of the arrow in Fig.~\ref{fig:R2TCd}(b)). The $S_v$ term in \eqref{eq:H123} drops out in the dual model~\footnote{The Hamiltonian $H$ and its dual $H_{\rm dual}$ will share the same phase diagram even without imposing the constraint, as the constraint used to reduce ${\cal H}_0$ to ${\cal H}'_0$ (${\cal H}_1$ to ${\cal H}'_1$) commute with the Hamiltonian $H_{\rm dual}(\lambda_1 , \lambda_2, \lambda_3 )$ ($H (\lambda_1 , \lambda_2, \lambda_3 )$).}. The four phases found in the dual Hamiltonian, labeled $P'_1$ through $P'_4$, are related to $P_1$ through $P_4$ phases by the duality in Fig.~\ref{fig:R2TCd}. Details of $P'_1$ to $P_4'$ phases can be found in Appendix~\ref{appendix:F}. The dual Hamiltonian is defined on a smaller Hilbert space, has \emph{local} order parameters to characterize phases as conventional SSB, and offers computational and conceptual simplicity over the original Hamiltonian.

\begin{figure}[ht]
\centering
\begin{overpic}[width=0.47\linewidth]{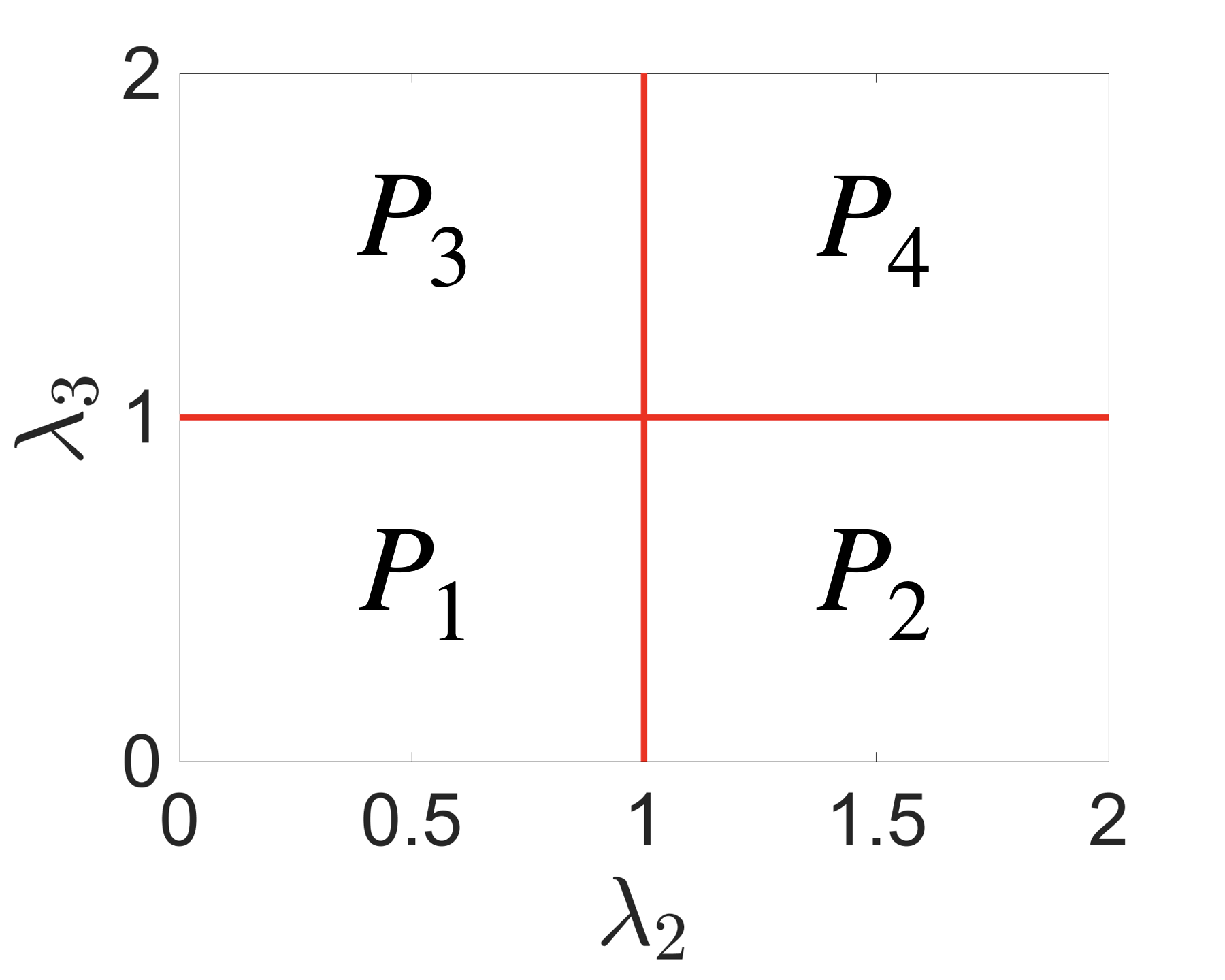}
\put(-5,78){(a)} 
\end{overpic}
\begin{overpic}[width=0.47\linewidth]{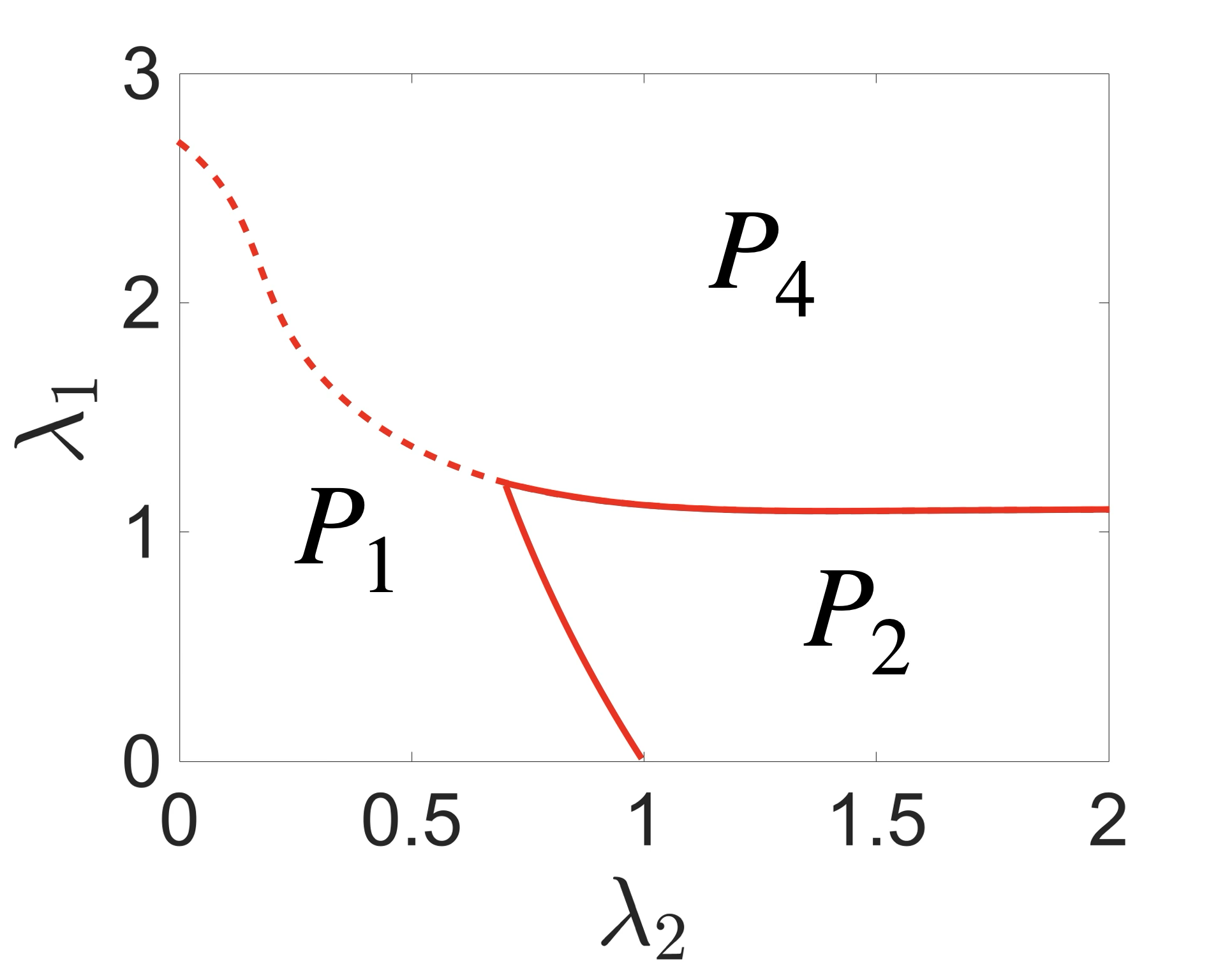}
\put(-5,78){(b)} 
\end{overpic}
\caption{Phase diagram of perturbed R2TC at (a) $\lambda_1 = 0$, and (b) $\lambda_3 = 0$. First- and second-order phase transitions are depicted by dotted and solid lines, respectively.}
\label{fig:Pd}
\end{figure}

Precise statements on the phase transitions can be made for the dual model at $\lambda_1 =0$, where it reduces to a collection of $\mathbb{Z}_N$ transverse Ising chains oriented along each horizontal and vertical line. The second order phase transitions are expected on the self-dual lines $\lambda_2 = 1$ and $\lambda_3 = 1$, with the phase diagram given exactly as in Fig.~\ref{fig:Pd}(a). For the case $\lambda_3=0$, we obtain the phase diagram shown in Fig.~\ref{fig:Pd}(b) by numerical calculation of the dual Hamiltonian with $\mathbb{Z}_2$ degrees of freedom. The phase transition between $P_1$ and $P_4$ is identified as a first-order transition (see Appendix~\ref{appendix:G}). To conclude, rather precise statements on the various transitions away from the dipolar topological state can be made by judicious use of the duality argument and numerical calculation.
\\

\section{Conclusion and Outlook}

We propose an extended framework for the simultaneous gauging of modulated symmetries, termed $n$-simultaneous gauging, which employs $n$ distinct gauge symmetry operators. In general, the sequence of simultaneous gauging and conventional gauging forms the most general framework for gauging modulated symmetries.

Our framework offers a route to engineer and probe exotic topological phases protected by modulated symmetries. In particular, simultaneous gauging~\footnote{Performed prior to Higgsing the matter sector.} yields an SPT phase with a mixed ’t Hooft anomaly between dipolar and higher-form symmetries. We expect the construction to generalize to broader classes of anomalies among generalized symmetries. A natural next step is to analyze higher-dimensional systems with richer higher-form or modulated symmetry structures, including non-Abelian groups.
\\

\section{Acknowledgment}

We are grateful to Zijian Song, Yun-Tak Oh, Chanbeen Lee, Seunghun Lee, Salvatore Pace, and Jacopo Gliozzi for helpful discussions. J.K. was supported by the education and training program of the
Quantum Information Research Support Center, funded through the National
research foundation of Korea(NRF) by the Ministry of science and ICT(MSIT) of
the Korean government(No. RS-2023-NR057243). J.Y.L. was supported by faculty startup grant at the University of Illinois, Urbana-Champaign. J.H.H. was supported by the National Research Foundation of Korea(NRF) grant funded by the Korea government(MSIT) (No. 2023R1A2C1002644). He acknowledges KITP, supported in part by the National Science Foundation under Grant No. NSF PHY-1748958.

\onecolumngrid
\appendix

\section{Sequential gauging of charge and dipole symmetries in (2+1)D}
\label{appendix:A}
In this section, we consider the $\mathbb{Z}_N$ charge and dipole symmetries, and carry out sequential gaugings in which each gauging process is governed by a single gauge symmetry operator.

\subsection{One charge symmetry and one dipole symmetry}

We first revisit the $\mathbb{Z}_N$ charge and $\mathbb{Z}_N$ dipole symmetries previously discussed in the context of 1-simultaneous gauging, defined as
\begin{align}
g_1&\equiv \prod_{\bf r} X_{\bf r}, & g_2&\equiv \prod_{\bf r} X_{\bf r}^x,
\end{align}
and assume the presence of translational symmetry. We now carry out sequential gaugings in which each gauging process is governed by a single gauge symmetry operator. One can first gauge the charge symmetry and then the dipole symmetry; however, the reverse order is not possible. The reason is that, under this restriction, $g_2$ must be written as $\prod_{\bf r} h_{\bf r}^x$, which by translational symmetry implies $g_1 = \prod_{\bf r} h_{\bf r}$. In this case, the procedure reduces to 1-simultaneous gauging of the two symmetries. Additionally, note that the local terms that commute with both $g_1$ and $g_2$ are $X_{\bf r}$, $Z_{{\bf r}-\hat{x}}Z_{{\bf r}}^{-2} Z_{{\bf r}+\hat{x}}$, $Z_{{\bf r}-\hat{x}}Z_{{\bf r}}^\dagger Z_{{\bf r}-\hat{x}-\hat{y}}^\dagger Z_{{\bf r}-\hat{y}}$, and $Z_{{\bf r}}^\dagger Z_{{\bf r}+\hat{y}}$.

Here, we focus on the topological phase that emerges after gauging. To this end, we start with the paramagnet Hamiltonian $H = -\sum_{\bf r} X_{\bf r} + {\rm h.c.}$ and track how it evolves under the gauging procedure and the imposition of the zero-flux condition.

% The resulting topological phase is then obtained by introducing appropriate stabilizers into the final Hamiltonian to enforce the zero-flux condition.

To gauge $g_1$ solely we will use the gauge symmetry operator $h_{\bf r}= X_{\bf r} \bar{Z}_{{\bf r}-\hat{y}/2}^\dagger \bar{Z}_{{\bf r}+\hat{y}/2} \bar{Z}_{{\bf r}-\hat{x}/2}^\dagger \bar{Z}_{{\bf r}+\hat{x}/2}$, which satisfies $g_1=\prod_{\bf r} h_{\bf r}$. The corresponding duality transformation is summarized as:
\begin{align}
X_{\bf r}&\rightarrow \bar{Z}_{{\bf r}-\hat{y}/2} \bar{Z}_{{\bf r}+\hat{y}/2}^\dagger \bar{Z}_{{\bf r}-\hat{x}/2} \bar{Z}_{{\bf r}+\hat{x}/2}^\dagger\nn
Z_{{\bf r}-\hat{x}}Z_{{\bf r}}^{-2} Z_{{\bf r}+\hat{x}} &\rightarrow \bar{X}_{{\bf r}-\hat{x}/2} \bar{X}_{{\bf r}+\hat{x}/2}^\dagger\nn
Z_{{\bf r}-\hat{x}}Z_{{\bf r}}^\dagger Z_{{\bf r}-\hat{x}-\hat{y}}^\dagger Z_{{\bf r}-\hat{y}} &\rightarrow \bar{X}_{{\bf r}-\hat{x}/2} \bar{X}_{{\bf r}-\hat{x}/2-\hat{y}}^\dagger\nn
Z_{{\bf r}}^\dagger Z_{{\bf r}+\hat{y}} &\rightarrow \bar{X}_{{\bf r}+\hat{y}/2}^\dagger,
\label{eq:duality5}
\end{align}
Therefore, the transformed Hamiltonian with zero-flux condition will be
\begin{align}
\bar{H}=-\sum_{\bf r} \bar{Z}_{{\bf r}-\hat{y}/2} \bar{Z}_{{\bf r} +\hat{y}/2}^\dagger \bar{Z}_{{\bf r}-\hat{x}/2} \bar{Z}_{{\bf r}+\hat{x}/2}^\dagger-\sum_{\bf r} \bar{X}_{{\bf r}+\hat{x}/2} \bar{X}_{{\bf r}+\hat{x}+\hat{y}/2} \bar{X}_{{\bf r}+\hat{x}/2+\hat{y}}^\dagger \bar{X}_{{\bf r}+\hat{y}/2}^\dagger+{\rm h.c.},
\end{align}
which is the $\mathbb{Z}_N$ toric code. In the dual lattice, the global symmetry is $\bar{g}_1=\prod_{\bf r}\bar{Z}_{{\bf r}+\hat{x}/2}$, where $\bar{g}_1$ can be obtained by adopting \eqref{eq:duality5} to $g_3$.

To gauge $\bar{g}_3$, the gauge symmetry operator which will be used is $\bar{h}_{{\bf r}+\hat{x}/2}=\bar{Z}_{{\bf r}+\hat{x}/2} \tilde{Z}_{{\bf r}+\hat{x}/2+\hat{y}/2} \tilde{Z}_{{\bf r}+\hat{x}/2-\hat{y}/2}^\dagger \tilde{Z}_{{\bf r}} \tilde{Z}_{{\bf r}+\hat{x}}^\dagger$, and implying $\bar{h}_{{\bf r}}=1$ leads to
\begin{align}
\bar{Z}_{{\bf r}-\hat{y}/2} \bar{Z}_{{\bf r}+\hat{y}/2}^\dagger \bar{Z}_{{\bf r}-\hat{x}/2} \bar{Z}_{{\bf r}+\hat{x}/2}^\dagger &\rightarrow \tilde{Z}_{{\bf r}-\hat{y}/2} \tilde{Z}_{{\bf r}+\hat{y}/2}^\dagger \tilde{Z}_{{\bf r}-\hat{x}/2+\hat{y}/2}^\dagger \tilde{Z}_{{\bf r}-\hat{x}/2-\hat{y}/2} \tilde{Z}_{{\bf r}-\hat{x}}^\dagger \tilde{Z}_{{\bf r}}^2 \tilde{Z}_{{\bf r}+\hat{x}}^\dagger%
\tilde{Z}_{{\bf r}+\hat{x}/2+\hat{y}/2} \tilde{Z}_{{\bf r}+\hat{x}/2-\hat{y}/2}^\dagger \nn
\bar{X}_{{\bf r}-\hat{x}/2} \bar{X}_{{\bf r}+\hat{x}/2}^\dagger &\rightarrow \tilde{X}_{{\bf r}}\nn
\bar{X}_{{\bf r}-\hat{x}/2} \bar{X}_{{\bf r}-\hat{x}/2-\hat{y}}^\dagger &\rightarrow \tilde{X}_{{\bf r}-\hat{x}/2-\hat{y}/2}\nn
\bar{X}_{{\bf r}+\hat{y}/2}^\dagger &\rightarrow \tilde{X}_{{\bf r}+\hat{y}/2}^\dagger.
\end{align}
Then, the transformed Hamiltonian with zero-flux condition will be 
\begin{align}
\tilde{H}&=-\sum_{\bf r} \tilde{Z}_{{\bf r}-\hat{y}/2} \tilde{Z}_{{\bf r}+\hat{y}/2}^\dagger \tilde{Z}_{{\bf r}-\hat{x}/2+\hat{y}/2}^\dagger \tilde{Z}_{{\bf r}-\hat{x}/2-\hat{y}/2} \tilde{Z}_{{\bf r}-\hat{x}}^\dagger \tilde{Z}_{{\bf r}}^2 %
\tilde{Z}_{{\bf r}+\hat{x}/2+\hat{y}/2} \tilde{Z}_{{\bf r}+\hat{x}/2-\hat{y}/2}^\dagger \tilde{Z}_{{\bf r}+\hat{x}}^\dagger\nn
&-\sum_{\bf r} \tilde{X}_{{\bf r}+\hat{x}/2+\hat{y}/2}^\dagger \bar{X}_{{\bf r}+\hat{x}+\hat{y}/2} \bar{X}_{{\bf r}+\hat{y}/2}^\dagger-\sum_{\bf r} X_{\bf r} X_{{\bf r}+\hat{x}/2+\hat{y}/2} X_{{\bf r} +\hat{y}}^\dagger X_{{\bf r}-\hat{x}/2+\hat{y}/2}^\dagger +{\rm h.c.}, \label{eq:tildeH}   
\end{align}
where there is one qudit at each vertex, one at each plaquette center, and one at each vertical edge. The overall duality can be summarized as:
\begin{align}
X_{\bf r}&\rightarrow \tilde{Z}_{{\bf r}-\hat{y}/2} \tilde{Z}_{{\bf r}+\hat{y}/2}^\dagger \tilde{Z}_{{\bf r}-\hat{x}/2+\hat{y}/2}^\dagger \tilde{Z}_{{\bf r}-\hat{x}/2-\hat{y}/2} \tilde{Z}_{{\bf r}-\hat{x}}^\dagger \tilde{Z}_{{\bf r}}^2 %
\tilde{Z}_{{\bf r}+\hat{x}/2+\hat{y}/2} \tilde{Z}_{{\bf r}+\hat{x}/2-\hat{y}/2}^\dagger \tilde{Z}_{{\bf r}+\hat{x}}^\dagger\nn
Z_{{\bf r}-\hat{x}}Z_{{\bf r}}^{-2} Z_{{\bf r}+\hat{x}} &\rightarrow \tilde{X}_{{\bf r}}\nn
Z_{{\bf r}-\hat{x}}Z_{{\bf r}}^\dagger Z_{{\bf r}-\hat{x}-\hat{y}}^\dagger Z_{{\bf r}-\hat{y}} &\rightarrow \tilde{X}_{{\bf r}-\hat{x}/2-\hat{y}/2}\nn
Z_{{\bf r}}^\dagger Z_{{\bf r}+\hat{y}} &\rightarrow \bar{X}_{{\bf r}+\hat{y}/2}^\dagger,
\end{align}
which can be rephrased as the 1-simultaneous gauging with the gauge symmetry operator $h_{\bf r}$ given as
\begin{align}
h_{\bf r}=X_{\bf r} \tilde{Z}_{{\bf r}-\hat{y}/2}^\dagger \tilde{Z}_{{\bf r}+\hat{y}/2} \tilde{Z}_{{\bf r}-\hat{x}/2+\hat{y}/2} \tilde{Z}_{{\bf r}-\hat{x}/2-\hat{y}/2}^\dagger \tilde{Z}_{{\bf r}-\hat{x}} \tilde{Z}_{{\bf r}}^{-2} %
\tilde{Z}_{{\bf r}+\hat{x}/2+\hat{y}/2}^\dagger \tilde{Z}_{{\bf r}+\hat{x}/2-\hat{y}/2} \tilde{Z}_{{\bf r}+\hat{x}}.
\end{align}

On the other hand, the Hamiltonian of anisotropic dipolar toric code is
\begin{align}
H=-\sum_{\bf r} Z_{{\bf r}-\frac{\hat{y}}{2}+\hat{x}} Z^{-2}_{{\bf r}-\frac{\hat{y}}{2}} Z_{{\bf r}-\frac{\hat{y}}{2}-\hat{x}} Z_{{\bf r}} Z^{-1}_{{\bf r}-\hat{y}} -\sum_{\bf r} X_{{\bf r}+\hat{x}} X^{-2}_{\bf r} X_{{\bf r}-\hat{x}}X_{{\bf r}+\frac{\hat{y}}{2}} X^{-1}_{{\bf r}-\frac{\hat{y}}{2}}+{\rm h.c.},\label{eq:atc}
\end{align}
where there is one qudit at each vertex, and one at each vertical edge. In fact, the anisotropic dipolar toric code emerges by projecting all plaquette qudits onto the $\bar{Z}=+1$ basis in the Hamiltonian given in \eqref{eq:tildeH}. From this perspective, the model obtained by sequential gauging differs from the anisotropic dipolar toric code. Additionally, one may say that 1-simultaneous gauging in (2+1)D may realize a gauging process that cannot be obtained through sequential gauging.

\subsection{Two charge symmetries and one dipole symmetry}

We now revisit the two $\mathbb{Z}_N$ charge and one $\mathbb{Z}_N$ dipole symmetries previously discussed in the context of 2-simultaneous gauging, defined as
\begin{align}
g_{e_{\rm h}}&\equiv \prod_{e_{\rm h}} X_{e_{\rm h}}, & g_{e_{\rm v}}&\equiv \prod_{e_{\rm v}} X_{e_{\rm v}}, & g_{\rm d}&\equiv \prod_{e_{\rm h}} X_{e_{\rm h}}^{-y_{e_{\rm h}}} \prod_{e_{\rm v}}X_{e_{\rm v}}^{x_{e_{\rm v}}},
\end{align}
and assume the presence of translational symmetry. We will examine the sequential gauging of charge and dipole symmetries, which will help clarify how 2-simultaneous gauging differs from sequential gauging in (2+1)D. As in the previous subsection, we begin by gauging the charge symmetry, followed by gauging the dipole symmetry, and starting with the Hamiltonian $H = -\sum_{e_{\rm h}} X_{e_{\rm h}} - \sum_{e_{\rm v}} X_{e_{\rm v}} + {\rm h.c.}$. The local terms that commute with these symmetries are $X_{e_{\rm h}}$, $X_{e_{\rm v}}$, $Z_{e_{\rm h}-\hat{x}}Z_{e_{\rm h}}^\dagger$, $Z_{e_{\rm v}-\hat{y}}Z_{e_{\rm v}}^\dagger$, $Z_{e_{\rm h}-\hat{y}}Z_{e_{\rm h}}^{-2} Z_{e_{\rm h}+\hat{y}}$, $Z_{e_{\rm v}-\hat{x}}Z_{e_{\rm v}}^{-2} Z_{e_{\rm v}+\hat{x}}$, $Z_{e_{\rm h}-\hat{y}}Z_{e_{\rm h}}^{-1} Z_{e_{\rm h}-\hat{x}-\hat{y}}^{-1} Z_{e_{\rm h}-\hat{x}}$, $Z_{e_{\rm v}-\hat{x}}Z_{e_{\rm v}}^\dagger Z_{e_{\rm v}-\hat{x}-\hat{y}}^\dagger Z_{e_{\rm v}-\hat{y}}$, and $Z_{e_{\rm h}-\hat{y}}Z_{e_{\rm h}}^\dagger Z_{e_{\rm h}-\hat{x}/2-\hat{y}/2} Z_{e_{\rm h}+\hat{x}/2-\hat{y}/2}^\dagger$.

To gauge $g_{e_{\rm h}}$ and $g_{e_{\rm v}}$, separately, the gauge symmetry operators $h_{e_{\rm h}}$ and $h_{e_{\rm v}}$ are:
\begin{align}
h_{e_{\rm h}}=X_{e_{\rm h}} \bar{Z}_{e_{\rm h}+\hat{y}/2} \bar{Z}_{e_{\rm h}-\hat{y}/2}^\dagger \bar{Z}_{e_{\rm h}+\hat{x}/2} \bar{Z}_{e_{\rm h}-\hat{x}/2}^\dagger\nn
h_{e_{\rm v}}=X_{e_{\rm v}} \tilde{Z}_{e_{\rm v}+\hat{y}/2}^\dagger \tilde{Z}_{e_{\rm v}-\hat{y}/2} \tilde{Z}_{e_{\rm v}+\hat{x}/2}^\dagger  \tilde{Z}_{e_{\rm v}-\hat{x}/2}.
\end{align}
The corresponding duality transformation is summarized as:
\begin{align}
X_{e_{\rm h}}&\rightarrow \bar{Z}_{e_{\rm h}+\hat{y}/2}^\dagger \bar{Z}_{e_{\rm h}-\hat{y}/2} \bar{Z}_{e_{\rm h}+\hat{x}/2}^\dagger \bar{Z}_{e_{\rm h}-\hat{x}/2}\nn
X_{e_{\rm v}}&\rightarrow \tilde{Z}_{e_{\rm v}+\hat{y}/2} \tilde{Z}_{e_{\rm v}-\hat{y}/2}^\dagger  \tilde{Z}_{e_{\rm v}+\hat{x}/2} \tilde{Z}_{e_{\rm v}-\hat{x}/2}^\dagger \nn
Z_{e_{\rm h}-\hat{x}}Z_{e_{\rm h}}^\dagger &\rightarrow \bar{X}_{e_{\rm h}-\hat{x}/2}\nn
Z_{e_{\rm v}-\hat{y}}Z_{e_{\rm v}}^\dagger &\rightarrow \tilde{X}_{e_{\rm v}-\hat{y}/2}^\dagger\nn
Z_{e_{\rm h}-\hat{y}}Z_{e_{\rm h}}^{-2} Z_{e_{\rm h}+\hat{y}} &\rightarrow \bar{X}_{e_{\rm h}-\hat{y}/2} \bar{X}_{e_{\rm h}+\hat{y}/2}^\dagger\nn
Z_{e_{\rm v}-\hat{x}}Z_{e_{\rm v}}^{-2} Z_{e_{\rm v}+\hat{x}} &\rightarrow \tilde{X}_{e_{\rm v}-\hat{x}/2}^\dagger \tilde{X}_{e_{\rm v}+\hat{x}/2}\nn
Z_{e_{\rm h}-\hat{y}}Z_{e_{\rm h}}^{-1} Z_{e_{\rm h}-\hat{x}-\hat{y}}^{-1} Z_{e_{\rm h}-\hat{x}} &\rightarrow \bar{X}_{e_{\rm h}-\hat{y}/2} \bar{X}_{e_{\rm h}-\hat{x}-\hat{y}/2}^\dagger\nn
Z_{e_{\rm v}-\hat{x}}Z_{e_{\rm v}}^\dagger Z_{e_{\rm v}-\hat{x}-\hat{y}}^\dagger Z_{e_{\rm v}-\hat{y}} &\rightarrow \tilde{X}_{e_{\rm v}-\hat{x}/2}^\dagger \tilde{X}_{e_{\rm v}-\hat{x}/2-\hat{y}}\nn
Z_{e_{\rm h}-\hat{y}}Z_{e_{\rm h}}^\dagger Z_{e_{\rm h}-\hat{x}/2-\hat{y}/2} Z_{e_{\rm h}+\hat{x}/2-\hat{y}/2}^\dagger &\rightarrow \bar{X}_{e_{\rm h}-\hat{y}/2} \tilde{X}_{e_{\rm h}-\hat{y}/2}^\dagger,
\end{align}
and the transformed Hamiltonian with zero-flux condition will be
\begin{align}
H'&=-\sum_{e_{\rm h}} \bar{Z}_{e_{\rm h}+\hat{y}/2}^\dagger \bar{Z}_{e_{\rm h}-\hat{y}/2} \bar{Z}_{e_{\rm h}+\hat{x}/2}^\dagger \bar{Z}_{e_{\rm h}-\hat{x}/2} - \sum_{e_{\rm v}} \tilde{Z}_{e_{\rm v}+\hat{y}/2} \tilde{Z}_{e_{\rm v}-\hat{y}/2}^\dagger \tilde{Z}_{e_{\rm v}+\hat{x}/2} \tilde{Z}_{e_{\rm v}-\hat{x}/2}^\dagger\nn
&-\sum_{e_{\rm h}}  \bar{X}_{e_{\rm h}-\hat{x}/2} \bar{X}_{e_{\rm h}-\hat{x}/2-\hat{y}}^\dagger \bar{X}_{e_{\rm h}-\hat{y}/2}^\dagger \bar{X}_{e_{\rm h}-\hat{x}-\hat{y}/2} -\sum_{e_{\rm v}} \tilde{X}_{e_{\rm v}-\hat{y}/2}^\dagger \tilde{X}_{e_{\rm v}-\hat{x}-\hat{y}/2} \tilde{X}_{e_{\rm v}-\hat{x}/2} \tilde{X}_{e_{\rm v}-\hat{x}/2-\hat{y}}^\dagger+ {\rm h.c.},
\end{align}
which corresponds to two copies of the toric code, with two qudits on the vertices and two qudits on the plaquette centers. The global symmetry in the dual lattice is $\bar{g}_1=\prod_{e_{\rm h}}\bar{Z}_{e_{\rm h}-\hat{y}/2} \prod_{e_{\rm v}} \tilde{Z}_{e_{\rm v}+\hat{x}/2}=\prod_p \bar{Z}_{p} \tilde{Z}_{p}$, where $p$ denotes plaquette centers.

To gauge $\bar{g}_1$, the gauge symmetry operator which will be used is $\bar{h}_{p}= \bar{Z}_{p} \tilde{Z}_{p} \check{Z}_{p+\hat{y}/2} \check{Z}_{p-\hat{y}/2}^\dagger \check{Z}_{p+\hat{x}/2} \check{Z}_{p-\hat{x}/2}^\dagger$~\footnote{One remark is that $\bar{g}_1$ can also be gauged by performing the 2-simultaneous gauging.}, and one possible choice of the corresponding duality transformation is summarized as:
\begin{align}
\bar{Z}_{e_{\rm h}+\hat{y}/2}^\dagger \bar{Z}_{e_{\rm h}-\hat{y}/2} \bar{Z}_{e_{\rm h}+\hat{x}/2}^\dagger \bar{Z}_{e_{\rm h}-\hat{x}/2} &\rightarrow \bar{Z}_{e_{\rm h}+\hat{y}/2}^\dagger \bar{Z}_{e_{\rm h}-\hat{y}/2}  \bar{Z}_{e_{\rm h}+\hat{x}/2}^\dagger \bar{Z}_{e_{\rm h}-\hat{x}/2} \check{Z}_{e_{\rm h}+\hat{y}} \check{Z}_{e_{\rm h}}^{-2} \check{Z}_{e_{\rm h}-\hat{y}}\nn
\tilde{Z}_{e_{\rm v}+\hat{y}/2} \tilde{Z}_{e_{\rm v}-\hat{y}/2}^\dagger  \tilde{Z}_{e_{\rm v}+\hat{x}/2} \tilde{Z}_{e_{\rm v}-\hat{x}/2}^\dagger &\rightarrow \tilde{Z}_{e_{\rm v}+\hat{y}/2} \tilde{Z}_{e_{\rm v}-\hat{y}/2}^\dagger \bar{Z}_{e_{\rm v}+\hat{x}/2}^\dagger \bar{Z}_{e_{\rm v}-\hat{x}/2} \check{Z}_{e_{\rm v}+\hat{x}}^\dagger \check{Z}_{e_{\rm v}}^2 \check{Z}_{e_{\rm v}-\hat{x}}^\dagger \nn
\bar{X}_{e_{\rm h}-\hat{x}/2} &\rightarrow \bar{X}_{e_{\rm h}-\hat{x}/2}\nn
\tilde{X}_{e_{\rm v}-\hat{y}/2}^\dagger &\rightarrow \tilde{X}_{e_{\rm v}-\hat{y}/2}^\dagger\nn
\bar{X}_{e_{\rm h}-\hat{y}/2} \bar{X}_{e_{\rm h}+\hat{y}/2}^\dagger &\rightarrow \check{X}_{e_{\rm h}}^\dagger \nn
\tilde{X}_{e_{\rm v}-\hat{x}/2}^\dagger \tilde{X}_{e_{\rm v}+\hat{x}/2} &\rightarrow \check{X}_{e_{\rm v}}\nn
\bar{X}_{e_{\rm h}-\hat{y}/2} \bar{X}_{e_{\rm h}-\hat{y}/2-\hat{x}}^\dagger &\rightarrow \bar{X}_{e_{\rm h}-\hat{y}/2} \bar{X}_{e_{\rm h}-\hat{x}-\hat{y}/2}^\dagger \check{X}_{e_{\rm h}-\hat{x}/2 -\hat{y}/2} \nn
\tilde{X}_{e_{\rm v}-\hat{x}/2}^\dagger \tilde{X}_{e_{\rm v}-\hat{x}/2-\hat{y}} &\rightarrow \bar{X}_{e_{\rm v}-\hat{x}/2} \bar{X}_{e_{\rm v}-\hat{x}/2-\hat{y}}^\dagger \check{X}_{e_{\rm v}-\hat{x}/2-\hat{y}/2}^\dagger\nn
\bar{X}_{e_{\rm h}-\hat{y}/2} \tilde{X}_{e_{\rm h}-\hat{y}/2}^\dagger &\rightarrow \bar{X}_{e_{\rm h}-\hat{y}/2}
\end{align}
Then, the transformed Hamiltonian with zero-flux condition will be
\begin{align}
H&=-\sum_{e_{\rm h}} \bar{Z}_{e_{\rm h}+\hat{y}/2}^\dagger \bar{Z}_{e_{\rm h}-\hat{y}/2}  \bar{Z}_{e_{\rm h}+\hat{x}/2}^\dagger \bar{Z}_{e_{\rm h}-\hat{x}/2} \check{Z}_{e_{\rm h}+\hat{y}} \check{Z}_{e_{\rm h}}^{-2} \check{Z}_{e_{\rm h}-\hat{y}} - \sum_{e_{\rm v}} \tilde{Z}_{e_{\rm v}+\hat{y}/2} \tilde{Z}_{e_{\rm v}-\hat{y}/2}^\dagger \bar{Z}_{e_{\rm v}+\hat{x}/2}^\dagger \bar{Z}_{e_{\rm v}-\hat{x}/2} \check{Z}_{e_{\rm v}+\hat{x}}^\dagger \check{Z}_{e_{\rm v}}^2 \check{Z}_{e_{\rm v}-\hat{x}}^\dagger\nn
&-\sum_{e_{\rm h}}  \bar{X}_{e_{\rm h}-\hat{x}/2} \bar{X}_{e_{\rm h}-\hat{x}/2-\hat{y}}^\dagger \bar{X}_{e_{\rm h}-\hat{y}/2}^\dagger \bar{X}_{e_{\rm h}-\hat{x}-\hat{y}/2} \check{X}_{e_{\rm h}-\hat{x}/2 -\hat{y}/2}^\dagger -\sum_{e_{\rm v}} \tilde{X}_{e_{\rm v}-\hat{y}/2}^\dagger \tilde{X}_{e_{\rm v}-\hat{x}-\hat{y}/2} \bar{X}_{e_{\rm v}-\hat{x}/2}^\dagger \bar{X}_{e_{\rm v}-\hat{x}/2-\hat{y}} \check{X}_{e_{\rm v}-\hat{x}/2-\hat{y}/2}^\dagger\nn
&-\sum_{e_{\rm h}} \check{X}_{e_{\rm h}}^\dagger \check{X}_{e_{\rm h}-\hat{x}} \bar{X}_{e_{\rm h}+\hat{y}/2} \bar{X}_{e_{\rm h}-\hat{x}+\hat{y}/2}^\dagger \check{X}_{e_{\rm h}-\hat{x}/2 +\hat{y}/2} \bar{X}_{e_{\rm h}-\hat{y}/2}^\dagger \bar{X}_{e_{\rm h}-\hat{x}-\hat{y}/2} \check{X}_{e_{\rm h}-\hat{x}/2 -\hat{y}/2}^\dagger+{\rm h.c.},\label{eq:tildeH2}
\end{align}
where two qudits reside on the vertices, one qudit on each plaquette, and one qudit on each edge. The overall duality can be summarized as:
\begin{align}
X_{e_{\rm h}}&\rightarrow \bar{Z}_{e_{\rm h}+\hat{y}/2}^\dagger \bar{Z}_{e_{\rm h}-\hat{y}/2}  \bar{Z}_{e_{\rm h}+\hat{x}/2}^\dagger \bar{Z}_{e_{\rm h}-\hat{x}/2} \check{Z}_{e_{\rm h}+\hat{y}} \check{Z}_{e_{\rm h}}^{-2} \check{Z}_{e_{\rm h}-\hat{y}}\nn
X_{e_{\rm v}}&\rightarrow \tilde{Z}_{e_{\rm v}+\hat{y}/2} \tilde{Z}_{e_{\rm v}-\hat{y}/2}^\dagger \bar{Z}_{e_{\rm v}+\hat{x}/2}^\dagger \bar{Z}_{e_{\rm v}-\hat{x}/2} \check{Z}_{e_{\rm v}+\hat{x}}^\dagger \check{Z}_{e_{\rm v}}^2 \check{Z}_{e_{\rm v}-\hat{x}}^\dagger \nn
Z_{e_{\rm h}-\hat{x}}Z_{e_{\rm h}}^\dagger &\rightarrow \bar{X}_{e_{\rm h}-\hat{x}/2}\nn
Z_{e_{\rm v}-\hat{y}}Z_{e_{\rm v}}^\dagger &\rightarrow \tilde{X}_{e_{\rm v}-\hat{y}/2}^\dagger\nn
Z_{e_{\rm h}-\hat{y}}Z_{e_{\rm h}}^{-2} Z_{e_{\rm h}+\hat{y}} &\rightarrow \check{X}_{e_{\rm h}}^\dagger\nn
Z_{e_{\rm v}-\hat{x}}Z_{e_{\rm v}}^{-2} Z_{e_{\rm v}+\hat{x}} &\rightarrow \check{X}_{e_{\rm v}}\nn
Z_{e_{\rm h}-\hat{y}}Z_{e_{\rm h}}^{-1} Z_{e_{\rm h}-\hat{x}-\hat{y}}^{-1} Z_{e_{\rm h}-\hat{x}} &\rightarrow \bar{X}_{e_{\rm h}-\hat{y}/2} \bar{X}_{e_{\rm h}-\hat{x}-\hat{y}/2}^\dagger \check{X}_{e_{\rm h}-\hat{x}/2 -\hat{y}/2}\nn
Z_{e_{\rm v}-\hat{x}}Z_{e_{\rm v}}^\dagger Z_{e_{\rm v}-\hat{x}-\hat{y}}^\dagger Z_{e_{\rm v}-\hat{y}} &\rightarrow \bar{X}_{e_{\rm v}-\hat{x}/2} \bar{X}_{e_{\rm v}-\hat{x}/2-\hat{y}}^\dagger \check{X}_{e_{\rm v}-\hat{x}/2-\hat{y}/2}^\dagger\nn
Z_{e_{\rm h}-\hat{y}}Z_{e_{\rm h}}^\dagger Z_{e_{\rm h}-\hat{x}/2-\hat{y}/2} Z_{e_{\rm h}+\hat{x}/2-\hat{y}/2}^\dagger &\rightarrow \bar{X}_{e_{\rm h}-\hat{y}/2},
\end{align}
which can be rephrased as the 2-simultaneous gauging with the gauge symmetry operators $h_{e_{\rm h}}$ and $h_{e_{\rm v}}$ given as
\begin{align}
h_{e_{\rm h}}&=X_{e_{\rm h}}\bar{Z}_{e_{\rm h}+\hat{y}/2} \bar{Z}_{e_{\rm h}-\hat{y}/2}^\dagger  \bar{Z}_{e_{\rm h}+\hat{x}/2} \bar{Z}_{e_{\rm h}-\hat{x}/2}^\dagger \check{Z}_{e_{\rm h}+\hat{y}}^\dagger \check{Z}_{e_{\rm h}}^{2} \check{Z}_{e_{\rm h}-\hat{y}}^\dagger\nn
h_{e_{\rm v}} &= X_{e_{\rm v}} \tilde{Z}_{e_{\rm v}+\hat{y}/2}^\dagger \tilde{Z}_{e_{\rm v}-\hat{y}/2} \bar{Z}_{e_{\rm v}+\hat{x}/2} \bar{Z}_{e_{\rm v}-\hat{x}/2}^\dagger \check{Z}_{e_{\rm v}+\hat{x}} \check{Z}_{e_{\rm v}}^{-2} \check{Z}_{e_{\rm v}-\hat{x}}.
\end{align}

In fact, the rank-2 toric code can be obtained by projecting all edge qudits onto the $\check{Z}=+1$ basis in the Hamiltonian given in \eqref{eq:tildeH2}. From this viewpoint, the model obtained through sequential gauging differs from the rank-2 toric code. Additionally, one may say that 2-simultaneous gauging in (2+1)D may realize a gauging process that cannot be obtained through sequential gauging.

\section{R2TC and its 1-form Symmetries}
\label{appendix:B}

In this section, we provide a comprehensive summary of rank-2 toric code (R2TC) and its 1-form symmetries~\cite{oh23}. The R2TC Hamiltonian is given as
\begin{align}
    H = -\sum_{v} S_{v}-\sum_{e_{\rm h}} S_{e_{\rm h}}-\sum_{e_{\rm v}} S_{e_{\rm v}}+ \textrm{h.c.}\label{eq:r2tc}
\end{align}
In this model, the mobility of anyons is constrained by the finite energy gap, and this restricted motion can be understood as a consequence of a dipole conservation of $\mathbb{Z}_N$ charge~\cite{oh22a,oh22b,oh23,kim23}. For simplicity, let us assume $N$ being prime.
To proceed, we define an anyon charge $c_v \in \mathbb{Z}_N$ for each vertex, which is associated with the value of the stabilizer $S_v = e^{i 2\pi c_v /N}$ which is defined in the main Fig. 1(a). This anyon charge obeys the dipole conservation laws represented as
\begin{align}
    &\sum_v x_v  c_v \equiv \sum_v y_v c_v \equiv 0 ~{\rm mod}~ N.
\end{align}
Consequently, the movement of a non-trivial anyon $c_v \neq 0$ is permitted only if $x_v$ or $y_v$ increases by $N$.

On the other hand, the charges $c_{e_{\rm h}}$ and $c_{e_{\rm v}}$ associated with the stabilizer $S_{e_{\rm h}}$ and $S_{e_{\rm v}}$, respectively, have the following dipole conservation law:
\begin{align}
   &\sum_{e_{\rm h}} y_{e_{\rm h}}c_{e_{\rm h}}+\sum_{e_{\rm v}} x_{e_{\rm v}} c_{e_{\rm v}} \equiv 0 ~{\rm mod}~ N. 
\end{align}
Therefore, $c_{e_{\rm h}}$ can move in increments of $N$ steps along the $y$-axis, while $c_{e_{\rm v}}$ can move in increments of $N$ steps along the $x$-axis. In other directions, $c_{e_{\rm h}}$ and $c_{e_{\rm v}}$ can move without constraints. 

% These mobility constraints also impose length restrictions on the 1-form symmetries.

There exist three 1-form symmetries, $g_{Z}^1$, $g_{Z}^2$, and $g_{Z}^3$, defined by a product of $Z$ operators along contractible or noncontractible loops, as well as three additional 1-form symmetries, $g_{X}^1$, $g_{X}^2$, and $g_{X}^3$, given by a product of $X$ operators. All of the six 1-form symmetries of R2TC act on vertices and plaquettes, but not on the edges where there are no quantum states.

\begin{figure}[ht]
\setlength{\abovecaptionskip}{5pt}
\centering
\begin{overpic}[width=0.38\columnwidth]{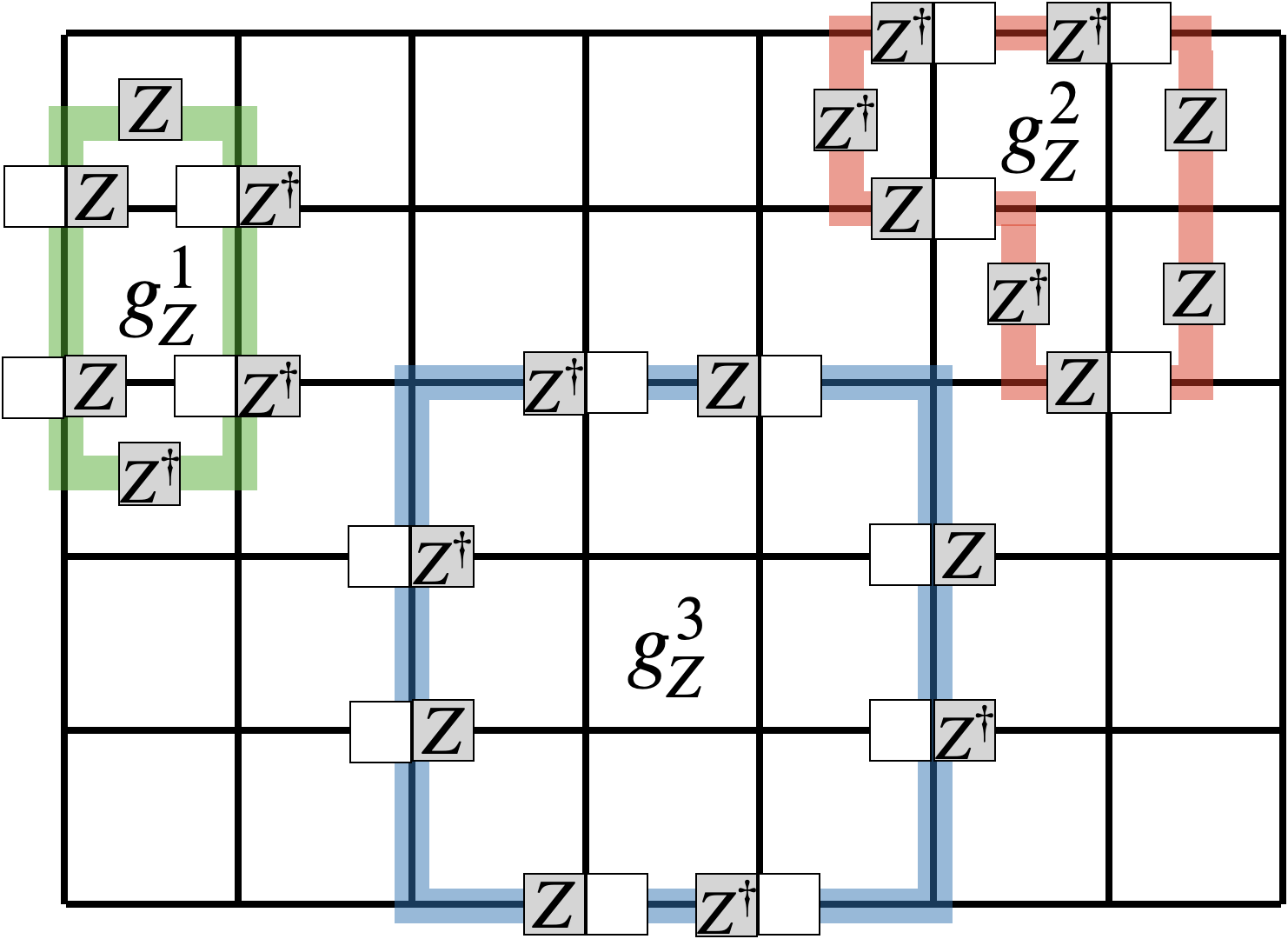}
\put(-5,112){(a)}
\end{overpic}
\begin{overpic}[width=0.43\columnwidth]{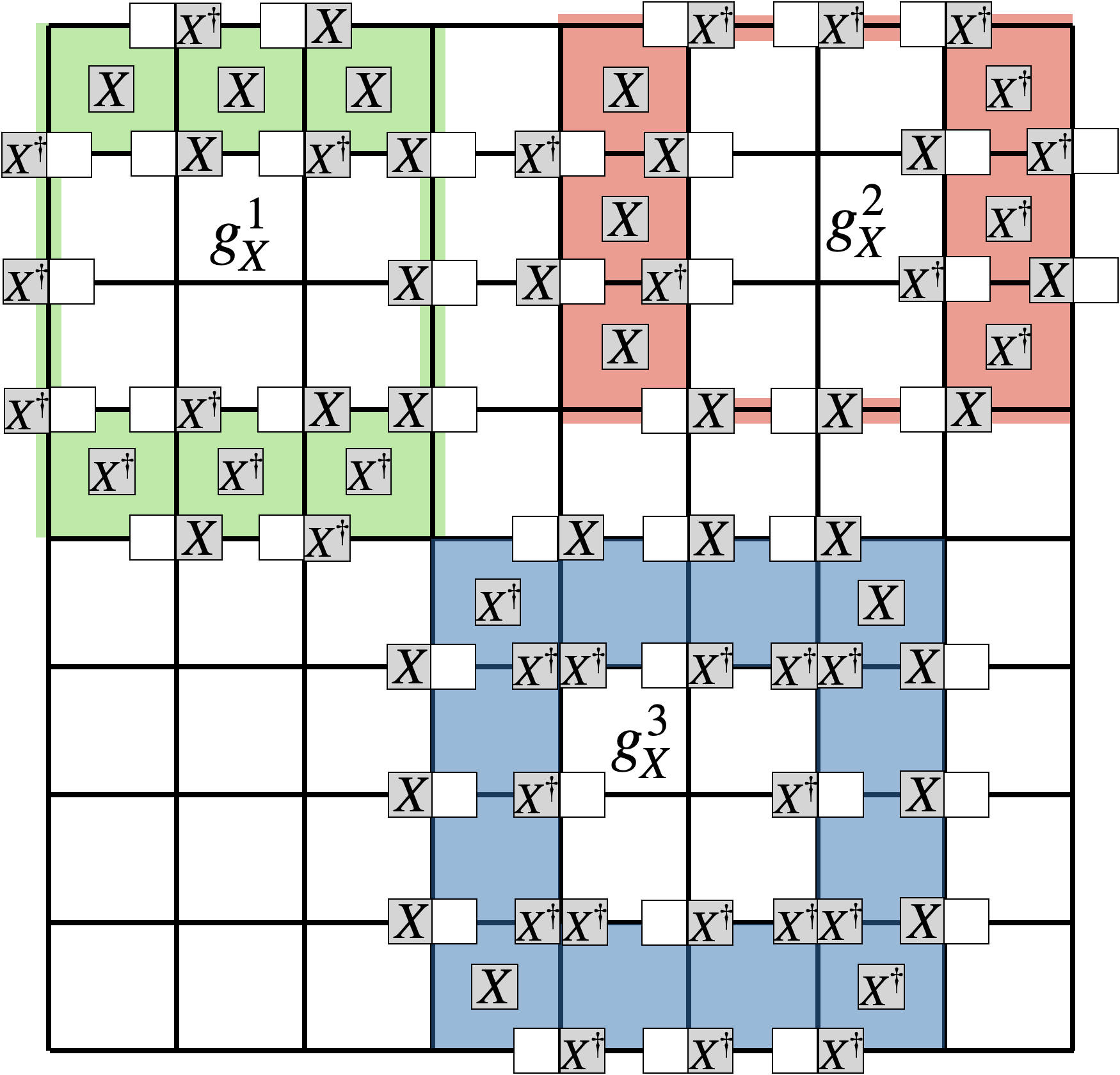}
\put(-5,100){(b)}
\end{overpic}

\caption{(a) Examples of the $g_Z^1$, $g_Z^2$, and $g_Z^3$ when $N=3$. (b) Examples of $g_X^1$, $g_X^2$, and $g_X^3$ when $N=3$.}
\label{fig:g123}
\end{figure}

\vspace{5pt} \noindent {\bf 1-form $Z$ symmetries.} 
\begin{itemize} 
    \item $g_Z^1$ is defined on the loop composed of horizontal lines that intersect the vertices of the dual lattice and vertical lines intersecting the vertices of the lattice. 
    \item $g_Z^2$ is defined on the loop composed of horizontal lines that intersect the vertices of the lattice and vertical lines that intersect the vertices of the dual lattice. 
    \item $g_Z^3$ is defined on a loop composed of the edges of the lattice, where the lengths of the horizontal and vertical lines are constrained to be integer multiples of $N$. Due to the length restriction, $g_Z^3$ is better characterized as a sublattice 1-form symmetry~\cite{oh23}.
\end{itemize}

For the non-contractible loop, the length restriction generally prevents $g_Z^3$ from being well-defined. However, it is possible for $(g_Z^3)^k$ to be well-defined when $L_x k \mod N = 0$ for horizontal loops and $L_y k \mod N = 0$ for vertical loops. This can be interpreted as $g_Z^3$ winding $k$ times around the torus, with a total length of $kL_x$ for horizontal loops and $kL_y$ for vertical loops. The smallest value of $k$ equals ${\rm lcm} (L_x,N)/N$ for horizontal loops and ${\rm lcm} (L_y,N)/N$ for vertical loops. Examples of $g_Z^1$, $g_Z^2$, and $g_Z^3$ for $N=3$ is depicted in Fig.~\ref{fig:g123}.

\vspace{5pt} \noindent {\bf 1-form $X$ symmetries.} 
\begin{itemize}
    \item $g_X^1$ is defined on the fattened loop composed of fattened horizontal lines that intersect the vertices of the dual lattice and vertical lines that intersect the vertices of the lattice, where the length of the fattened horizontal lines be a multiple of $N$. This length restriction implies that for the non-contractible loop along the $x$-axis, $(g_X^1)^k$ with $k = \mathrm{lcm}(L_x, N)/N$ is a well-defined symmetry. 
    \item $g_X^2$ is defined on the fattened loop composed of horizontal lines passing through the vertices of the lattice and fattened vertical lines intersecting the vertices of the dual lattice, where the length of the fattened vertical lines be a multiple of $N$. This length restriction implies that for the non-contractible loop along the $y$-axis, $(g_X^2)^k$ with $k = \mathrm{lcm}(L_y, N)/N$ is a well-defined symmetry. 
    \item $g_X^3$ is defined on the fattened loop composed of the fattened edges of the dual lattice. 
\end{itemize}
Examples of $g_X^1$, $g_X^2$, and $g_X^3$ for $N=3$ are depicted in Fig.~\ref{fig:g123}(b).

The 1-form symmetries defined on non-contractible loops are logical operators that act on the degenerate ground state manifold. As previously noted, while some 1-form symmetries may not be well-defined on non-contractible loops, their powers may be well-defined. As the number of logical operators depends on the lattice size, the ground state degeneracy also depends on the lattice size. We refer to the GSD analysis using logical operators in \cite{oh23}.

\begin{figure}[ht]
\includegraphics[width=0.42\columnwidth]{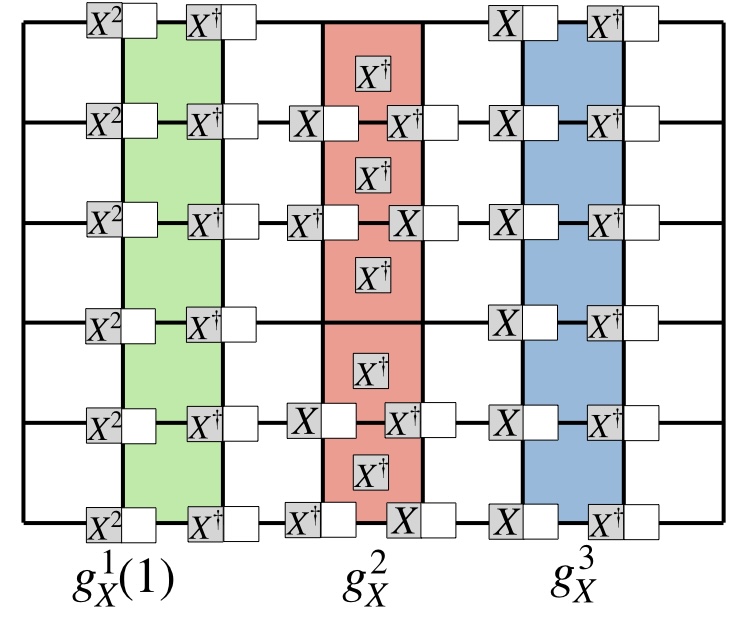}
\caption{$g_X^1(1)$, $g_X^2$, and $g_X^3$ defined along the $y$-axis with a smooth horizontal boundary for $N=3$}
\label{fig:x123}
\end{figure}

Finally, to aid in the understanding of (11) in the main text, Fig.~\ref{fig:x123} presents $g_X^1(1)$, $g_X^2$, and $g_X^3$ defined along the $y$-axis with a smooth horizontal boundary for $N=3$. Only $g_X^1(x_{v_l})$ is position-dependent and is generally expressed as
\begin{align}
g_X^1(x_{v_l})=\prod_{x_{v_l}} X_{v_l}(X_{v_l} X_{v_l}^\dagger)^{x_{v_{l}}}.
\end{align}

\section{Projection of dCS and anyon conservation laws}
\label{appendix:C}

After projecting all edge qudits in the dCS to the $X = +1$ eigenstate, remaining qudits at the vertices and plaquette centers form the ground state of R2TC. Indeed, the two stabilizers $S_{e_{\rm h}}'$ and $S_{e_{\rm v}}'$ commute with $X=+1$ measurement at the edges and reduce to $S_{e_{\rm h}}$ and $S_{e_{\rm v}}$ of R2TC, respectively. The collapsed state is an +1 eigenstate of $g_X^1$, $g_X^2$, and $g_X^3$, which are the symmetries of dCS that commute with the measurement. The remaining three stabilizers $S_{v_r}'$, $S_{p}'$, and $S_{v_l}'$ in Fig. 2(c) by themselves do not commute with the measurements at the edges, but a carefully chosen product of them:
\begin{figure}[H]
    \centering
    \includegraphics[width=0.2\linewidth]{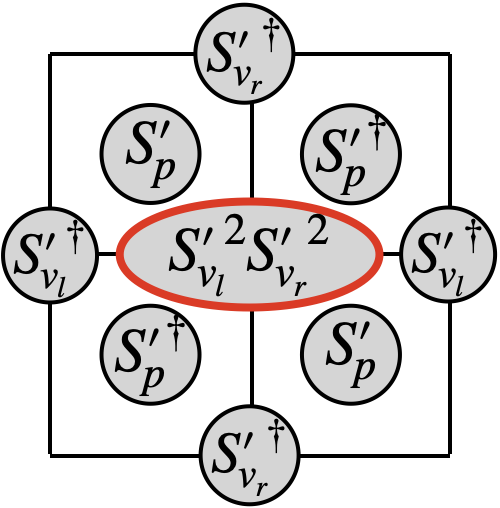}
    % \caption{Caption}
    % \label{fig:placeholder}
\end{figure}
\noindent which equals $S_v$ of the R2TC. The collapsed state is $S_v = +1$ eigenstate regardless of the outcome of the edge-$X$ measurement. Other ground states of R2TC can be generated by further applying $g_Z^1$ through $g_Z^3$ on noncontractible loops to the collapsed state.

The symmetries $g_{\rm d}$, $g_{e_{\rm h}}$, and $g_{e_{\rm v}}$ that are gauged during the gauging procedure are closely related to the conservation laws of the R2TC. Invoking the relation $h_{e_{\rm h}} = X_{e_{\rm h}} S_{e_{\rm h}}$ and $h_{e_{\rm v}} = X_{e_{\rm v}} S_{e_{\rm v}}$, and the fact that the $g_{\rm d}$, $g_{e_{\rm h}}$, and $g_{e_{\rm v}}$ become +1, it follows that
\begin{align}
&\prod_{e_{\rm h}} S_{e_{\rm h}}=1, ~~~~~~~\prod_{e_{\rm v}} S_{e_{\rm v}}=1\nn
&\Bigl(\prod_{e_{\rm h}} (S_{e_{\rm h}})^{-y_{e_{\rm h}}} \prod_{e_{\rm v}} (S_{e_{\rm v}})^{x_{e_{\rm v}}}\Bigl )^{N/\gcd (L_x,L_y,N)}=1 , 
\label{eq:conservation}
\end{align}
in the post-measurement state. The first two relations signify the conservation of two types of anyon charge and the third equation implies total dipole conservation. In summary, the conservation of anyon charge and dipole moment in the topological model emerging after gauging is a direct consequence of the symmetries that are subjected to the gauging procedure.

\section{Dipole bundle symmetry}
\label{appendix:D}

In this section, we provide a rigorous definition of the dipole bundle symmetry in two dimensions. The concept of the bundle symmetry was introduced in an earlier work on the dipolar SPT chain in one dimension~\cite{dSPT}. The dipole bundle symmetry of the dCS, defined on a patch $A_\alpha$ enclosed by a contractible loop and denoted as $g_b(A_\alpha)$, is expressed as:
\begin{align}
g_{\rm b}(A_\alpha) &:= \prod_{e_{\rm h} \in A_\alpha} X_{e_{\rm h}}^{-y_{e_{\rm h}}} \prod_{e_{\rm v} \in A_\alpha} X_{e_{\rm v}}^{x_{e_{\rm v}}}\nn
&= \prod_{e_{\rm h}\in A_\alpha} (S_{e_{\rm h}}')^{-y_{e_{\rm h}}}\prod_{e_{\rm v}\in A_\alpha} (S_{e_{\rm v}}')^{x_{e_{\rm v}}}.
% \label{eq:dipole-symmetry-of-dCS}
\end{align}
The lattice, which can be defined under both open and periodic boundary conditions, is covered by multiple overlapping patches enclosed by contractible loops, satisfying $A_\alpha \cap A_{\beta} \neq \emptyset$ and $\bigcup_\alpha A_\alpha = \text{lattice}$. For a given patch $A_\alpha$, $g_{\rm b}(A_\alpha)$ commutes with stabilizers that are fully supported inside $A_\alpha$, but not with those defined across its boundary. When the lattice is defined under open boundary condition, $A_{\alpha}$ can encompass the entire lattice and yield $g_{\rm b}(\text{lattice}) = g_{\rm d}$.

This is a notable departure from conventional global symmetry and highlights that the dCS under open and periodic boundary condition are protected by the same symmetries. As mentioned in the main text, $g_{\rm D}$, defined on the torus, commutes with the dCS Hamiltonian; however, its expression varies with lattice size. Arguing that $g_{\rm D}$ is the symmetry that protects the state is inappropriate, as the short-range entangled nature of the dCS—characterized by the reduced density matrix or correlation functions—is independent of the lattice size. Moreover, $g_{\rm d}$ does not commute with the dCS Hamiltonian in general. Therefore, the dCS is most appropriately described as being protected by the dipole bundle symmetry $g_{\rm b}(A_\alpha)$.

\section{Analysis of Boundary modes of dCS}
\label{appendix:E}

We will re-examine the degeneracy of the boundary modes by constructing a symmetric boundary Hamiltonian. Let $U$ be the product of $\textrm{CZ}$ and $\textrm{CZ}^\dag$ operations that transform a paramagnetic state into the dCS state for a lattice with smooth boundary, and define $\overline{O}\,{:=}\,U O U^\dagger$. In this formalism, we can write $g_{\rm h}^{u} = \prod_{e_{\rm h}} \overline{X}_{e_{\rm h}}$ and  $g_{\rm v}^{\rm u} = \prod_{v_l \in u} \overline{Z}_{v_l}$ and $g_{\rm d}^{\rm u} = \prod_{v_l \in u} \overline{Z}_{v_l}^{x_{v_l}}$~\footnote{$g_{\rm d}^{\rm u} \rightarrow ( g_{\rm d}^{\rm u} )^{N/\gcd(L_x,N)}$ for periodic boundary condition}. One can show that the simplest symmetry-allowed local terms are
\begin{align} \overline{Z}_{e_{\rm h}} \overline{Z}_{e_{\rm h}+\hat{x}}^\dagger, ~~\overline{X}_{v_l -\hat{x}} (\overline{X}_{v_l }^{\dagger})^2 \overline{X}_{v_l + \hat{x}}
\end{align}
and their conjugates. The boundary Hamiltonian $ \sum_{e_{\rm h}} \overline{Z}_{e_{\rm h}} \overline{Z}_{e_{\rm h}+\hat{x}}^\dagger + {\rm h.c.}$ on horizontal edge qubits spontaneously breaks $g_{\rm h}^u$, resulting in an $N$-fold degeneracy. Similarly, the Hamiltonian $\sum_{v_l } \overline{X}_{v_l -\hat{x}} (\overline{X}_{v_l }^{\dagger})^2 \overline{X}_{v_l + \hat{x}} + {\rm h.c.}$ breaks $g_{\rm v}^u$ and $g_{\rm d}^u$ ($(g_{\rm d}^u)^{N/\gcd(L_x,N)}$) with $N^2$ ($N\gcd(L_x,N)$) degeneracy. Transverse field terms such as $\overline{X}_{e_{\rm h}}$ and $\overline{Z}_{v_l}$ that could lift the degeneracy at the boundary are forbidden by symmetry.

Now we consider a rough horizontal boundary. Under the similar process, the symmetry-allowed local boundary operators for a rough horizontal boundary are $\overline{X}_p \overline{X}_{p+\hat{x}}^\dagger$ and $\overline{Z}_{e_{\rm v}-\hat{x}}(\overline{Z}_{e_{\rm v}}^\dagger)^2 \overline{Z}_{e_{\rm v}+\hat{x}}$. These two terms commute and can coexist on the same boundary. The interaction term $\overline{X}_p \overline{X}_{p+\hat{x}}^\dagger$ induces SSB of the $\mathbb{Z}_N$ symmetry $g_{\rm h}$. Conversely, the interaction $\overline{Z}_{e_{\rm v}-\hat{x}}(\overline{Z}_{e_{\rm v}}^\dagger)^2 \overline{Z}_{e_{\rm v}+\hat{x}}$ leads to SSB of the $\mathbb{Z}_{N}\times \mathbb{Z}_N$ ($\mathbb{Z}_N \times \mathbb{Z}_{\gcd(L_x,N)}$) symmetry associated with $g_{\rm v}$ and $g_{\rm d}$ ($g_{\rm d}^{N/\gcd(L_x,N)}$). Thus, the SSB characteristics for both smooth and rough horizontal boundaries are identical.

\section{Characteristics of the phases of $H$ and $H_{\rm dual}$}
\label{appendix:F}

The model $H(\lambda_1, \lambda_2, \lambda_3)$ hosts four distinct phases $P_1$ through $P_4$. The $P_1$ is the regime in which the ground state remains in the same phase as R2TC, and occurs for small values of $\lambda_\alpha$, with $\alpha = 1, 2, 3$. The $P_2$ ($P_3$) phase, realized for $\lambda_2 \gg 1$ ($\lambda_3 \gg 1$), has finite average of $X_{v_l}$ ($X_{v_r}$) and the concomitant condensation of the quasiparticles associated with $S_{e_{\rm v}}$ ($S_{e_{\rm h}}$) along with the confinement of quasiparticles associated with $S_{v}$. The $P_2$ ($P_3$) phase is partially topological, as evidenced by the deconfinement of $S_{e_{\rm h}}$ ($S_{e_{\rm v}}$) quasiparticles and the existence of the corresponding Wilson loop operators $g_X^2$ ($g_X^1$). In the $P_4$ phase, both $S_{e_{\rm h}}$ and $S_{e_{\rm v}}$ quasiparticles are condensed and $S_{v}$ quasiparticles are confined, with finite expectation values of $X_p$, $X_{v_l}$, and $X_{v_r}$. This phase emerges under (i) $\lambda_1 \gg 1$ with $\lambda_2, \lambda_3 > 0$, or (ii) $\lambda_2, \lambda_3 \gg 1$ with $\lambda_1 > 0$. The ground state is unique.

The four phases found in the dual Hamiltonian, labeled $P'_1$ through $P'_4$, are related to $P_1$ through $P_4$ phases by the $K$ duality. The $P_1'$ phase ($\lambda_\alpha \ll 1$) with a unique ground state is characterized by finite averages $\langle X_{e_{\rm h}} \rangle$ and $\langle X_{e_{\rm v}} \rangle$, and represents the dual of the dipolar topological phase of R2TC. The $P'_2$ phase ($\lambda_1, \lambda_3 \ll 1 \ll \lambda_2$) exhibits ferromagnetic order with finite $\langle Z_{e_{\rm v}}\rangle$ along each vertical chain for the $e_{\rm v}$ qudits while the $e_{\rm h}$ qudits remain  paramagnetic with finite $\langle X_{e_{\rm h}} \rangle$. There is a sub-extensive GSD of $N^{L_x}$ with the factor $N$ arising from SSB of the subsystem symmetry $s_1'=\prod_{e_{\rm v} \in C^{\rm v}} X_{e_{\rm v}}$ on each vertical chain. The $P'_3$ phase ($\lambda_1, \lambda_2 \ll 1 \ll \lambda_3$) is the $P'_2$ phase with $e_{\rm v}$ and $e_{\rm h}$ qudits interchanged. The $P'_4$ phase ($\lambda_1 \gg 1$, $\lambda_2, \lambda_3>0$ or $\lambda_2, \lambda_3 \gg 1$, $\lambda_1>0$) is understood as the SSB of two charge symmetries $g_{\rm v}$ and $g_{\rm h}$ and the dipole symmetry $g_{\rm D}$; it is characterized by finite $\langle Z_{e_{\rm h}} \rangle$ and $\langle Z_{e_{\rm v}} \rangle$. The dual Hamiltonian is defined on a smaller Hilbert space, has \emph{local} order parameters to characterize phases as conventional SSB, and offers computational and conceptual simplicity over the original Hamiltonian.

Most of the phase characteristics have been previously studied or are well understood, except for the $P_2$ and $P_3$ phases. Therefore, we will analyze the $P_2$ phase Hamiltonian under the periodic boundary condition at the fixed point of $P_2$, where $\lambda_1, \lambda_3 = 0$ and $\lambda_2 \to \infty$. At the fixed point, $S_{e_v}$ is suppressed to zero, and the remaining terms in the Hamiltonian acquire the local symmetry $l_{e_{\rm v}}$.
\begin{figure}[H]
\centering
\includegraphics[width=0.24\linewidth]{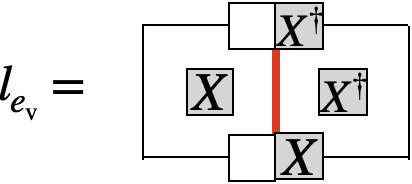},
\end{figure}
\noindent and two subsystem symmetries $s_1=\prod_{p \in C^{\rm v}} Z_p$ and $s_2=(\prod_{v \in C^{\rm v}} Z_{v_r}^{x_v})^{N/\gcd(L_x,N)}$ along an arbitrary vertical loop $C^{\rm v}$~\footnote{Under the open boundary condition, the expression of $s_2$ will be changed to $s_2=\prod_{v \in C^{\rm v}} Z_{v_r}^{x_v}$}.

We now turn to the calculation of the ground state degeneracy (GSD). We will use two distinct approaches to calculate the GSD: (i) counting the number of independent stabilizer identities, and (ii) counting the number of independent operations that generate distinct ground states.

To implement the first method, the underlying logical framework is as follows. At the fixed point, the ground state corresponds to the +1 eigenstate of the operators $(X_{v_l}, S_v, S_{e_{\rm h}})$. Given that $X_{v_l} = 1$ for all $v_l$ qudits, the effective Hamiltonian can be described in terms of the stabilizers $S_v^{P_2}$ and $S_{e_{\rm h}}$, where $S_v^{P_2}$ is expressed as
\begin{figure}[H]
    \centering
    \includegraphics[width=0.25\linewidth]{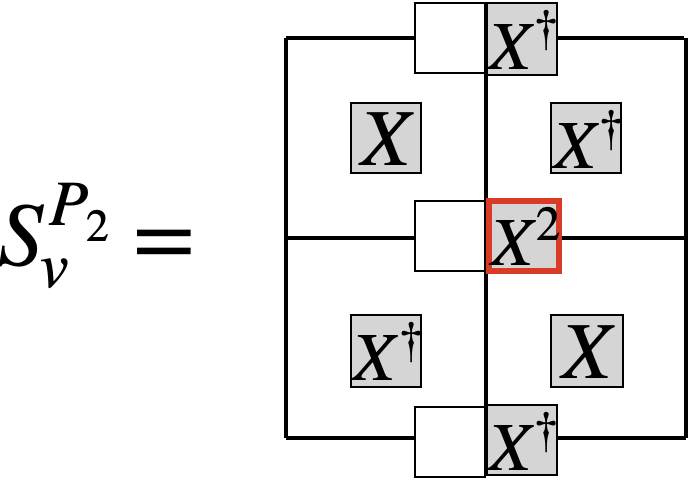}.
\end{figure}
\noindent This formulation is applicable within the reduced Hilbert space $\{ | h_{v_r}, h_p \rangle \}$. The stabilizer $S_{e_{\rm h}}$ corresponds to the plaquette term of the $\mathbb{Z}_{N}$ toric code. Consequently, it satisfies a single identity, expressed as $\prod_{e_{\rm h}} S_{e_{\rm h}} = 1$. The identities associated with $S_v^{P_2}$ are
\begin{align}
\prod_{v \in C^{\rm v}} S_v^{P_2}=1, ~~~~~~~~~\prod_{v \in C^{\rm v}} (S_v^{{P_2}})^{y_vN/\gcd(L_y,N)}=1,
\end{align}
where $C^{\rm v}$ is a vertical line. Therefore the GSD of the $P_2$ phase Hamiltonian is $N^{1+L_x}\gcd(L_y, N)$.

To implement the second method, we first examine the ground state of the $P_2$ phase Hamiltonian, which is characterized as the +1 eigenstate of $S_{v}^{P_2}$'s:
\begin{align}
\ket{\psi}=\prod_{e_{\rm h}} (1+S_{e_{\rm h}}+S_{e_{\rm h}}^2+\cdots+S_{e_{\rm h}}^{N-1})\ket{\overline{0}},
\end{align}
where $\ket{\overline{0}}$ is a +1 eigenstate of $X$ for every qudit. There are four distinct operations that yield different ground states. Two of these operations, $\prod_{p \in C^{\rm h}} Z_p$ and $\prod_{v_r \in C^{\rm v}} Z_{v_r}$, are effectively analogous to the Wilson loop operators in the toric code. Here, $C^{\rm h}$ refers the horizontal line. The contributions of these operations to the GSD are calculated to be $N$ and $N$, respectively. The remaining two operations are $\prod_{p \in C^{\rm v}} Z_p$ and $\prod_{v_l \in C^{\rm v}} Z_{v_l}^{y_{v_l}N/\gcd(L_y,N)}$. Notably, two operators of the form $\prod_{v_l \in C^{\rm v}} Z_{v_l}^{y_{v_l}N/\gcd(L_y,N)}$, defined on different vertical lines, are related through the product of $S_{e_{\rm h}}$ and $\prod_{p \in C^{\rm v}} Z_p$. Consequently, a straightforward counting of the contributions to the GSD from these two operations yields factors of $N^{L_x}$ and $\gcd(L_y, N)$, respectively. However, not all of $\prod_{p \in C^{\rm v}} Z_p$ operations are independent. First of all, the product of $S_{e_{\rm h}}$, $\prod_{e_{\rm h}} S_{e_{\rm h}}^{-y_{{e_{\rm h}}}N/\gcd(L_y,N)}$ is equivalent to $\prod_p Z_p^{N/\gcd(L_y,N)}$, which means $\prod_p Z_p^{N/\gcd(L_y,N)}$ is not the operation that give different ground state. Therefore, $N^{L_x}$ should be divided by $\gcd(L_y,N)$. Furthermore, consider the $\prod_{p} Z_p$, which can be obtained by the product of $\prod_{p \in C^{\rm v}} Z_p$. This operation is equivalent to $\prod_{p \in C^{\rm h}} Z_p^{L_y}$, using the $S_{e_{\rm h}}$ operations. In number theory, it is well known that the least value of $nL_y \mod N$ is $\gcd(L_y, N)$, where $n$ is a natural number. This means that when $\gcd(L_y, N) = 1$, the operation $\prod_{p \in C^{\rm h}} Z_p$, which was considered to contribute $N$ to the ground state, is in fact fully derivable from other operations and does not contribute at all. Therefore, in general, we can conclude that $\prod_{p \in C^{\rm h}} Z_p$ contributes $\gcd(L_y, N)$ to the GSD. Collecting all contributions, this leads to $N^{1+L_x} \gcd(L_y, N)$, which matches the result obtained using the first method.

For the last, we consider another ground state of the $P_2$ phase Hamiltonian, which is characterized as the +1 eigenstate of $S_{e_{\rm h}}$'s:
\begin{align}
\ket{\psi}=\prod_{v} (1+S_v^{P_2}+(S_v^{P_2})^2+\cdots + (S_v^{P_2})^{N-1})\ket{0},
\end{align}
and try the similar analysis, where $\ket{0}$ is +1 eigenstate of every $Z$. There exists a local symmetry denoted by $l_{e_{\rm v}}$, where $l_{e_{\rm v}}$ and $l_{e_{\rm v}-\hat{y}}$ represent equivalent operations connected by $S_{v}^{P_2}$. Consequently, for each vertical line, there is one operation of $l_{e_{\rm v}}$. Additionally, another identity arises from the relation $\prod_{e_{\rm v} \in C^{\rm h}} l_{e_{\rm v}}^{L_y}$ is effectively equivalent to $\prod_{e_{\rm h}} l_{e_{\rm v}} = 1$. Leveraging number theory, we deduce that $\prod_{e_{\rm v} \in C^{\rm h}} l_{e_{\rm v}}^{\gcd(L_y, N)} = 1$, implying that the contribution of the local symmetry is $N^{L_x-1}\gcd(L_y, N)$. The remaining two symmetries, $\prod_{p \in C^{\rm v}} X_p$ and $\prod_{v_r \in C^{\rm h}} Z_{v_r}$, are analogous to Wilson loop operators in the toric code. Their contributions are $N$ and $N$, respectively, as the local symmetry connecting these symmetries across different loops has been fully accounted for. Thus, the GSD is determined to be $N^{1+L_x} \gcd(L_y, N)$ once more.

The properties of the phases of $H(\lambda_1, \lambda_2, \lambda_3)$ and $H_{\rm dual}(\lambda_1, \lambda_2, \lambda_3)$ are summarized in Table \ref{tab:P}.

\begin{table}[H]
\centering
    \begin{tabular}{c|c|c}
        \toprule
        \hline
        ~Phase~ & GSD & Order parameter operators\\
        \hline
        $P_1$& ~$N^3 c_x c_y c_{x,y}$~ & ~$g_Z^1$, $g_Z^2$, $g_Z^3$, $g_X^1$, $g_X^2$, $g_X^3$ (NC)~\\
        \hline
        $P_2$& $N^{1+L_x}c_y$ & $g_Z^1$, $g_X^2$ (NC)\\
        \hline
        $P_3$& $N^{1+L_y}c_x$ & $g_Z^2$, $g_X^1$ (NC)\\
        \hline
        $P_4$& $1$ & $X_p$, $X_{v_{l}}$, $X_{v_{r}}$\\
        \hline
        \hline
        $P_1'$& $1$ & $X_{e_{\rm h}}$, $X_{e_{\rm v}}$\\
        \hline
        $P_2'$& $N^{L_x}$ & $X_{e_{\rm h}}$, $Z_{e_{\rm v}}$\\
        \hline
        $P_3'$& $N^{L_y}$ & $Z_{e_{\rm h}}$, $X_{e_{\rm v}}$\\
        \hline
        $P_4'$& $N^2 c_{x,y}$& $Z_{e_{\rm h}}$, $Z_{e_{\rm v}}$\\
        \hline
        \bottomrule
    \end{tabular}
    \caption{The characteristics of the phases $P_1$ through $P_4$ and $P_1'$ through $P_4'$. $c_x$, $c_y$, and $c_{x,y}$ denote $\gcd(L_x,N)$, $\gcd(L_y,N)$, and $\gcd(L_x,L_y,N)$, respectively. NC refers to operators defined on noncontractible loops.}
    \label{tab:P}
\end{table}

\section{Details of Numerical calculations}
\label{appendix:G}

\begin{figure}[ht]
\centering
\includegraphics[width=0.2\linewidth]{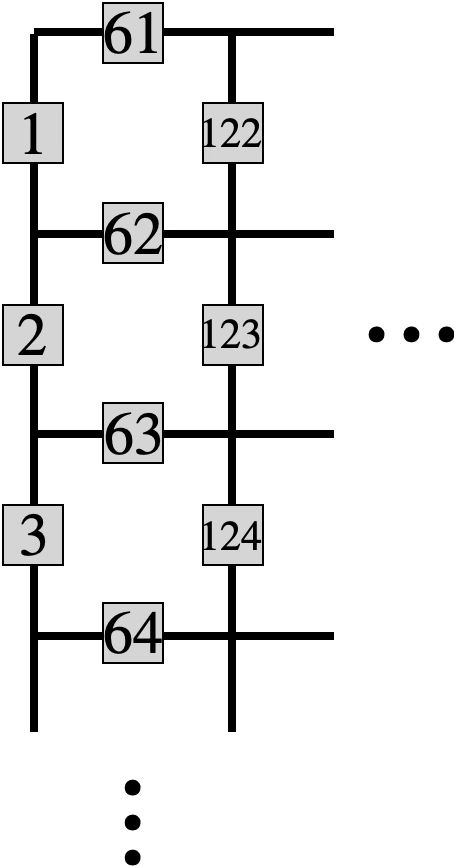}
\caption{Labeling of the qudits of a lattice structure with dimensions $L_x=10$ and $L_y=60$. The numbering follows a column-major order, with labels increasing sequentially as you move downward within each column of qudits before progressing to the next column of qudits.}
\label{fig:label}
\end{figure}

The numerical calculations of $H_{\rm dual}$ with $\mathbb{Z}_{2}$ degrees of freedom were performed employing a quasi-1D implementation of the density matrix renormalization group algorithm using matrix product states. We selected a lattice structure with dimensions $L_x = 10$ and $L_y = 60$, comprising $2$ qudits per unit cell under open boundary condition. The qudits were labeled as illustrated in Fig.~\ref{fig:label}. Key parameters included a cutoff of $10^{-9}$, a maximum bond dimension of $50$, and an energy tolerance of $10^{-9}$. The choice of $L_y = 60$ was sufficient to attain the critical point within the range $1<\lambda_2 <1.1$ under the condition $\lambda_1 = \lambda_3 = 0$. Since $L_y=60$ is already considerably large to execute the quasi-1D calculation, and given the exponential growth of the bond dimension with increasing $L_y$, which is inherent to the quasi-1D implementation, obtaining the exact result becomes computationally challenging. Therefore, we selected $L_x = 10$ to ensure computational feasibility. As a potential avenue for future research, one could explore the use of projected entangled pair states to achieve a more precise determination of the phase diagram.

\begin{figure}[ht]
\centering
\begin{overpic}[width=0.32\linewidth]{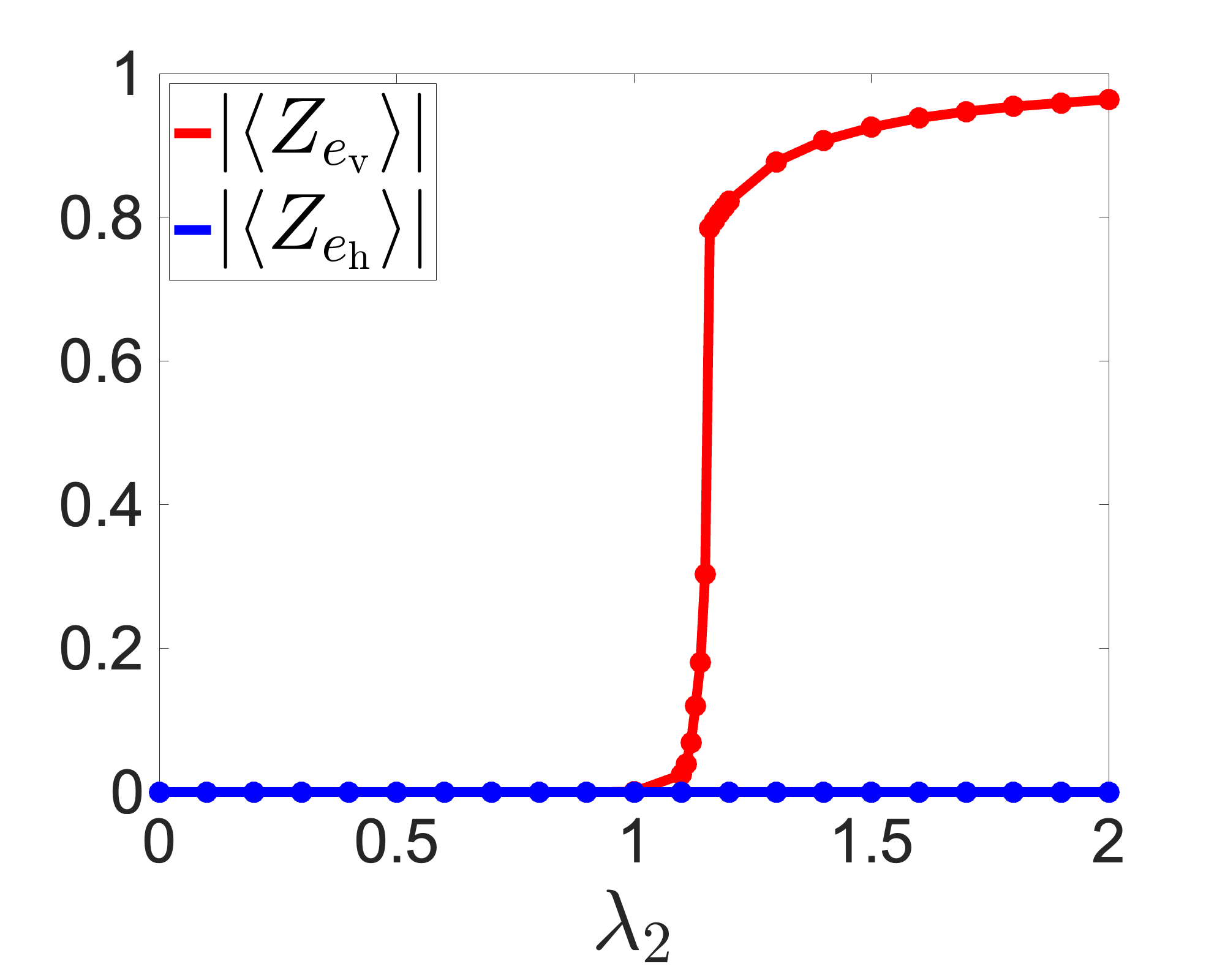}
\put(-3,75){(a)}
\end{overpic}
\begin{overpic}[width=0.32\linewidth]{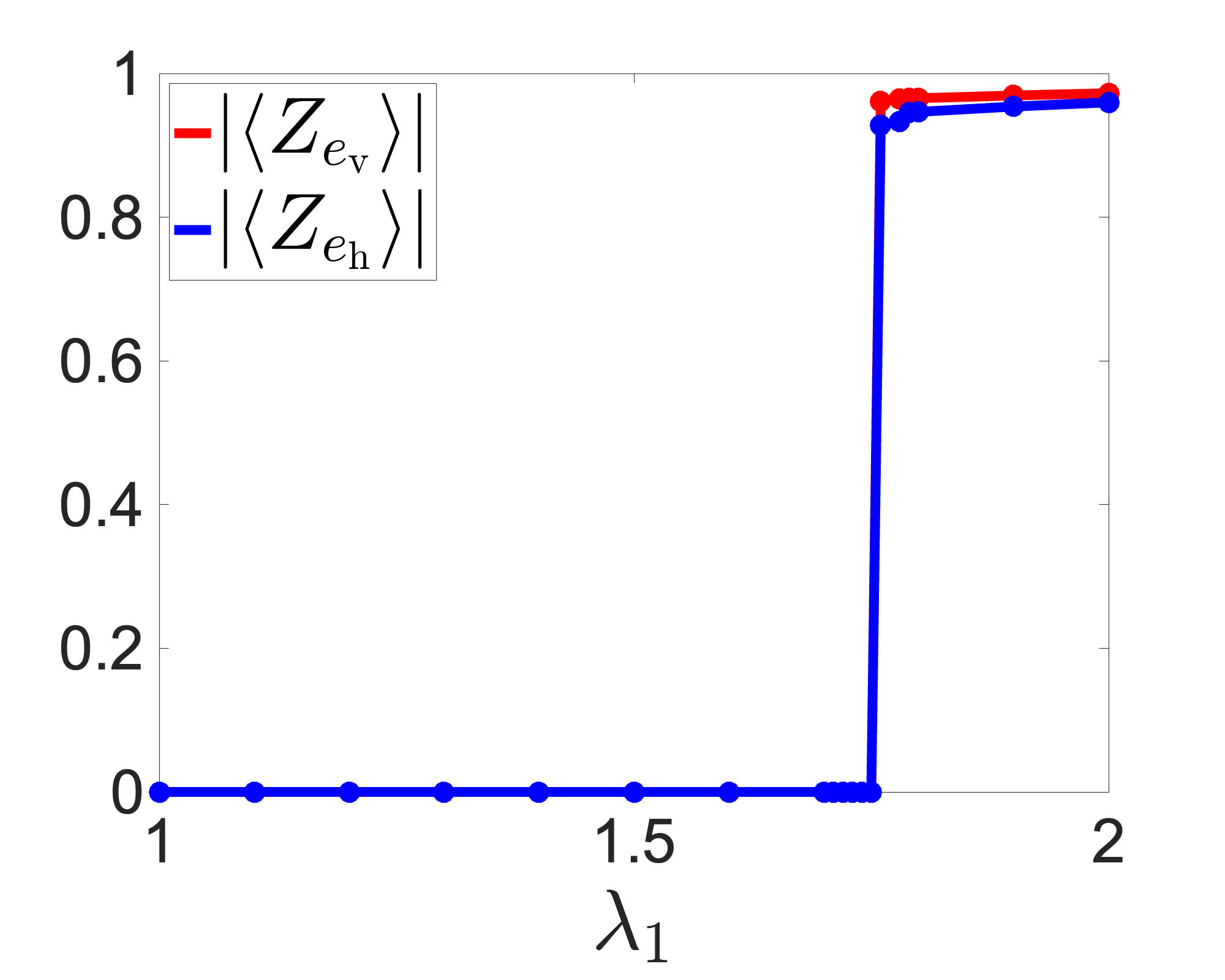}
\put(-3,75){(b)}
\end{overpic}
\begin{overpic}[width=0.32\linewidth]{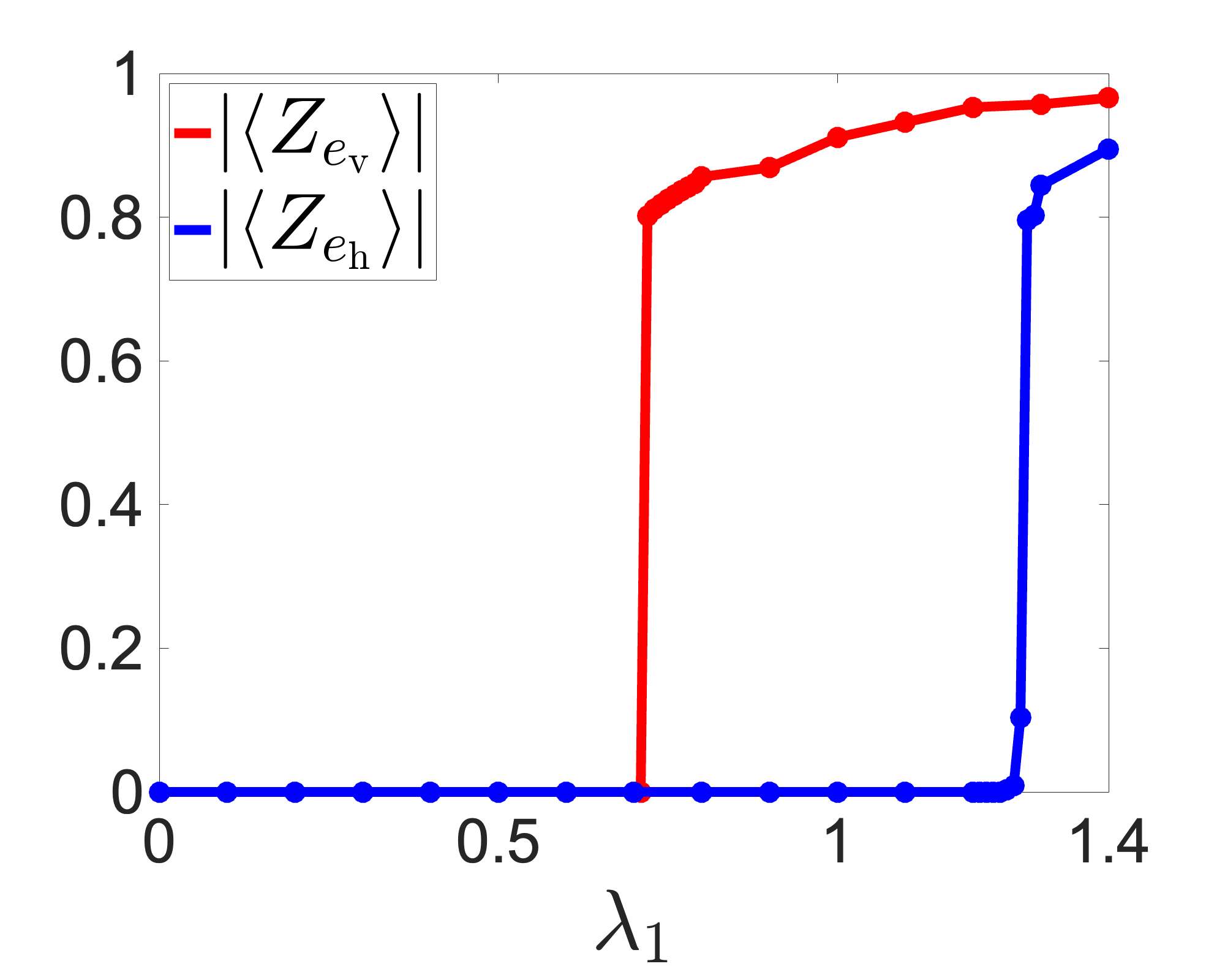}
\put(-3,75){(c)}
\end{overpic}  
\caption{The graphs of the absolute values of $\langle Z_{e_{\rm h}} \rangle$ and $\langle Z_{e_{\rm v}} \rangle$ are presented for the following cases: (a) $\lambda_1 = \lambda_3 = 0$, (b) $\lambda_2 = 0.3, \lambda_3 = 0$, and (c) $\lambda_2 = 0.8, \lambda_3 = 0$. The dots represent the values obtained from the numerical simulations.}
\label{fig:num_cal}
\end{figure}

Additionally, we present graphs derived from the numerical calculations of the expectation values of $Z_{e_{\rm h}}$ and $Z_{e_{\rm v}}$, aimed at identifying the critical points for various values of $\lambda_1$, $\lambda_2$, and $\lambda_3$. We show three specific cases: $\lambda_1 = \lambda_3 = 0$ [Fig.~\ref{fig:num_cal}(a)], $\lambda_2 = 0.3, \lambda_3 = 0$ [Fig.~\ref{fig:num_cal}(b)], and $\lambda_2 = 0.8, \lambda_3 = 0$ [Fig.~\ref{fig:num_cal}(c)], which illustrate the phase transitions between $P_1$ and $P_3$, $P_1$ and $P_4$, and $P_1$, $P_2$, and $P_4$, respectively. Based on the numerical results, we anticipate that the phase transition between $P_1$ and $P_4$ is of first order, whereas the other transitions correspond to the second order phase transitions. The source codes and the numerical results are available at \cite{kim25zenodo}.

\twocolumngrid

\bibliography{ref}

@article{han24,
  title = {Dipolar background field theory and dipolar braiding statistics},
  author = {Han, Jung Hoon},
  journal = {Phys. Rev. B},
  volume = {109},
  issue = {23},
  pages = {235127},
  numpages = {8},
  year = {2024},
  month = {Jun},
  publisher = {American Physical Society},
  doi = {10.1103/PhysRevB.109.235127},
  url = {https://link.aps.org/doi/10.1103/PhysRevB.109.235127}
}

@Article{huang23,
	title={{A Chern-Simons theory for dipole symmetry}},
	author={Xiaoyang Huang},
	journal={SciPost Phys.},
	volume={15},
	pages={153},
	year={2023},
	publisher={SciPost},
	doi={10.21468/SciPostPhys.15.4.153},
	url={https://scipost.org/10.21468/SciPostPhys.15.4.153},
}

@article{lake22,
	title = {Dipolar Bose-Hubbard model},
	author = {Lake, Ethan and Hermele, Michael and Senthil, T.},
	journal = {Phys. Rev. B},
	volume = {106},
	issue = {6},
	pages = {064511},
	numpages = {14},
	year = {2022},
	month = {Aug},
	publisher = {American Physical Society},
	doi = {10.1103/PhysRevB.106.064511},
	url = {https://link.aps.org/doi/10.1103/PhysRevB.106.064511}
}

@article{sala22,
  title = {Dynamics in Systems with Modulated Symmetries},
  author = {Sala, Pablo and Lehmann, Julius and Rakovszky, Tibor and Pollmann, Frank},
  journal = {Phys. Rev. Lett.},
  volume = {129},
  issue = {17},
  pages = {170601},
  numpages = {7},
  year = {2022},
  month = {Oct},
  publisher = {American Physical Society},
  doi = {10.1103/PhysRevLett.129.170601},
  url = {https://link.aps.org/doi/10.1103/PhysRevLett.129.170601}
}

@article{gromov24,
  title = {Colloquium: Fracton matter},
  author = {Gromov, Andrey and Radzihovsky, Leo},
  journal = {Rev. Mod. Phys.},
  volume = {96},
  issue = {1},
  pages = {011001},
  numpages = {26},
  year = {2024},
  month = {Jan},
  publisher = {American Physical Society},
  doi = {10.1103/RevModPhys.96.011001},
  url = {https://link.aps.org/doi/10.1103/RevModPhys.96.011001}
}

@article{ma18,
  title = {Fracton topological order from the Higgs and partial-confinement mechanisms of rank-two gauge theory},
  author = {Ma, Han and Hermele, Michael and Chen, Xie},
  journal = {Phys. Rev. B},
  volume = {98},
  issue = {3},
  pages = {035111},
  numpages = {15},
  year = {2018},
  month = {Jul},
  publisher = {American Physical Society},
  doi = {10.1103/PhysRevB.98.035111},
  url = {https://link.aps.org/doi/10.1103/PhysRevB.98.035111}
}

@article{bulmash18,
  title = {Higgs mechanism in higher-rank symmetric U(1) gauge theories},
  author = {Bulmash, Daniel and Barkeshli, Maissam},
  journal = {Phys. Rev. B},
  volume = {97},
  issue = {23},
  pages = {235112},
  numpages = {25},
  year = {2018},
  month = {Jun},
  publisher = {American Physical Society},
  doi = {10.1103/PhysRevB.97.235112},
  url = {https://link.aps.org/doi/10.1103/PhysRevB.97.235112}
}

@article{nandkishore-fracton-review,
author = {Nandkishore, Rahul M. and Hermele, Michael},
title = {Fractons},
journal = {Annual Review of Condensed Matter Physics},
volume = {10},
number = {1},
pages = {295-313},
year = {2019},
doi = {10.1146/annurev-conmatphys-031218-013604},
URL = {https://doi.org/10.1146/annurev-conmatphys-031218-013604},
eprint = {https://doi.org/10.1146/annurev-conmatphys-031218-013604}
}

@Article{delfino23,
	title={{Effective fractonic behavior in a two-dimensional exactly solvable spin  liquid}},
	author={Guilherme Delfino and Weslei Bernardino Fontana and Pedro R. S. Gomes and Claudio Chamon},
	journal={SciPost Phys.},
	volume={14},
	pages={002},
	year={2023},
	publisher={SciPost},
	doi={10.21468/SciPostPhys.14.1.002},
	url={https://scipost.org/10.21468/SciPostPhys.14.1.002}
}

@article{ebisu24,
  title = {Multipole and fracton topological order via gauging foliated symmetry protected topological phases},
  author = {Ebisu, Hiromi and Honda, Masazumi and Nakanishi, Taiichi},
  journal = {Phys. Rev. Res.},
  volume = {6},
  issue = {2},
  pages = {023166},
  numpages = {18},
  year = {2024},
  month = {May},
  publisher = {American Physical Society},
  doi = {10.1103/PhysRevResearch.6.023166},
  url = {https://link.aps.org/doi/10.1103/PhysRevResearch.6.023166}
}

@article{gorantla23,
  title = {Gapped lineon and fracton models on graphs},
  author = {Gorantla, Pranay and Lam, Ho Tat and Seiberg, Nathan and Shao, Shu-Heng},
  journal = {Phys. Rev. B},
  volume = {107},
  issue = {12},
  pages = {125121},
  numpages = {22},
  year = {2023},
  month = {Mar},
  publisher = {American Physical Society},
  doi = {10.1103/PhysRevB.107.125121},
  url = {https://link.aps.org/doi/10.1103/PhysRevB.107.125121}
}

@misc{delfino23b,
      title={2D Fractons from Gauging Exponential Symmetries}, 
      author={Guilherme Delfino and Claudio Chamon and Yizhi You},
      year={2023},
      eprint={2306.17121},
      archivePrefix={arXiv},
      primaryClass={cond-mat.str-el}
}

@article{ebisu23a,
  title = {Symmetric higher rank topological phases on generic graphs},
  author = {Ebisu, Hiromi},
  journal = {Phys. Rev. B},
  volume = {107},
  issue = {12},
  pages = {125154},
  numpages = {14},
  year = {2023},
  month = {Mar},
  publisher = {American Physical Society},
  doi = {10.1103/PhysRevB.107.125154},
  url = {https://link.aps.org/doi/10.1103/PhysRevB.107.125154}
}

@article{watanabe23,
    author = {Watanabe, Haruki and Cheng, Meng and Fuji, Yohei},
    title = "{Ground state degeneracy on torus in a family of $Z_N$ toric code}",
    journal = {Journal of Mathematical Physics},
    volume = {64},
    number = {5},
    pages = {051901},
    year = {2023},
    month = {05},
    issn = {0022-2488},
    doi = {10.1063/5.0134010},
    url = {https://doi.org/10.1063/5.0134010}
}

@article{zechmann23,
  title = {Fractonic Luttinger liquids and supersolids in a constrained Bose-Hubbard model},
  author = {Zechmann, Philip and Altman, Ehud and Knap, Michael and Feldmeier, Johannes},
  journal = {Phys. Rev. B},
  volume = {107},
  issue = {19},
  pages = {195131},
  numpages = {16},
  year = {2023},
  month = {May},
  publisher = {American Physical Society},
  doi = {10.1103/PhysRevB.107.195131},
  url = {https://link.aps.org/doi/10.1103/PhysRevB.107.195131}
}

@article{lake23,
  title = {Dipole condensates in tilted Bose-Hubbard chains},
  author = {Lake, Ethan and Lee, Hyun-Yong and Han, Jung Hoon and Senthil, T.},
  journal = {Phys. Rev. B},
  volume = {107},
  issue = {19},
  pages = {195132},
  numpages = {20},
  year = {2023},
  month = {May},
  publisher = {American Physical Society},
  doi = {10.1103/PhysRevB.107.195132},
  url = {https://link.aps.org/doi/10.1103/PhysRevB.107.195132}
}

@article{dSPT,
  title = {Topological quantum chains protected by dipolar and other modulated symmetries},
  author = {Han, Jung Hoon and Lake, Ethan and Lam, Ho Tat and Verresen, Ruben and You, Yizhi},
  journal = {Phys. Rev. B},
  volume = {109},
  issue = {12},
  pages = {125121},
  numpages = {19},
  year = {2024},
  month = {Mar},
  publisher = {American Physical Society},
  doi = {10.1103/PhysRevB.109.125121},
  url = {https://link.aps.org/doi/10.1103/PhysRevB.109.125121}
}

@article{ro24,
  title = {Scaling and Localization in Multipole-Conserving Diffusion},
  author = {Han, Jung Hoon and Lake, Ethan and Ro, Sunghan},
  journal = {Phys. Rev. Lett.},
  volume = {132},
  issue = {13},
  pages = {137102},
  numpages = {6},
  year = {2024},
  month = {Mar},
  publisher = {American Physical Society},
  doi = {10.1103/PhysRevLett.132.137102},
  url = {https://link.aps.org/doi/10.1103/PhysRevLett.132.137102}
}

@ARTICLE{surowka22,
AUTHOR={Grosvenor, Kevin T.  and Hoyos, Carlos  and Peña-Benítez, Francisco  and Surówka, Piotr },
TITLE={Space-Dependent Symmetries and Fractons},
JOURNAL={Frontiers in Physics},
VOLUME={9},
YEAR={2022},
URL={https://www.frontiersin.org/journals/physics/articles/10.3389/fphy.2021.792621},
DOI={10.3389/fphy.2021.792621},
ISSN={2296-424X}
}

@article{feldmeier20,
  title = {Anomalous Diffusion in Dipole- and Higher-Moment-Conserving Systems},
  author = {Feldmeier, Johannes and Sala, Pablo and De Tomasi, Giuseppe and Pollmann, Frank and Knap, Michael},
  journal = {Phys. Rev. Lett.},
  volume = {125},
  issue = {24},
  pages = {245303},
  numpages = {6},
  year = {2020},
  month = {Dec},
  publisher = {American Physical Society},
  doi = {10.1103/PhysRevLett.125.245303},
  url = {https://link.aps.org/doi/10.1103/PhysRevLett.125.245303}
}

@article{gromov20,
  title = {Fracton hydrodynamics},
  author = {Gromov, Andrey and Lucas, Andrew and Nandkishore, Rahul M.},
  journal = {Phys. Rev. Res.},
  volume = {2},
  issue = {3},
  pages = {033124},
  numpages = {11},
  year = {2020},
  month = {Jul},
  publisher = {American Physical Society},
  doi = {10.1103/PhysRevResearch.2.033124},
  url = {https://link.aps.org/doi/10.1103/PhysRevResearch.2.033124}
}

@article{morningstar20,
	title={Kinetically constrained freezing transition in a dipole-conserving system},
	author={Morningstar, Alan and Khemani, Vedika and Huse, David A},
	journal={Physical Review B},
	volume={101},
	number={21},
	pages={214205},
	year={2020},
	publisher={APS}
}

@article{sala20,
	title={Ergodicity breaking arising from Hilbert space fragmentation in dipole-conserving Hamiltonians},
	author={Sala, Pablo and Rakovszky, Tibor and Verresen, Ruben and Knap, Michael and Pollmann, Frank},
	journal={Physical Review X},
	volume={10},
	number={1},
	pages={011047},
	year={2020},
	publisher={APS}
}

@article{sachdev02,
	title={Mott insulators in strong electric fields},
	author={Sachdev, Subir and Sengupta, K and Girvin, SM},
	journal={Physical Review B},
	volume={66},
	number={7},
	pages={075128},
	year={2002},
	publisher={APS}
}

@article{skinner23,
	title = {Exact solution for the filling-induced thermalization transition in a one-dimensional fracton system},
	author = {Pozderac, Calvin and Speck, Steven and Feng, Xiaozhou and Huse, David A. and Skinner, Brian},
	journal = {Phys. Rev. B},
	volume = {107},
	issue = {4},
	pages = {045137},
	numpages = {15},
	year = {2023},
	month = {Jan},
	publisher = {American Physical Society},
	doi = {10.1103/PhysRevB.107.045137},
	url = {https://link.aps.org/doi/10.1103/PhysRevB.107.045137}
}

@article{gorantla22,
	title = {Global dipole symmetry, compact Lifshitz theory, tensor gauge theory, and fractons},
	author = {Gorantla, Pranay and Lam, Ho Tat and Seiberg, Nathan and Shao, Shu-Heng},
	journal = {Phys. Rev. B},
	volume = {106},
	issue = {4},
	pages = {045112},
	numpages = {37},
	year = {2022},
	month = {Jul},
	publisher = {American Physical Society},
	doi = {10.1103/PhysRevB.106.045112},
	url = {https://link.aps.org/doi/10.1103/PhysRevB.106.045112}
}

@article{yizhiyou20,
author = {Pretko, Michael and Chen, Xie and You, Yizhi},
title = {Fracton phases of matter},
journal = {International Journal of Modern Physics A},
volume = {35},
number = {06},
pages = {2030003},
year = {2020},
doi = {10.1142/S0217751X20300033},
URL = {https://doi.org/10.1142/S0217751X20300033},
eprint = {https://doi.org/10.1142/S0217751X20300033}
}

@article{bakr20,
  title = {Subdiffusion and Heat Transport in a Tilted Two-Dimensional Fermi-Hubbard System},
  author = {Guardado-Sanchez, Elmer and Morningstar, Alan and Spar, Benjamin M. and Brown, Peter T. and Huse, David A. and Bakr, Waseem S.},
  journal = {Phys. Rev. X},
  volume = {10},
  issue = {1},
  pages = {011042},
  numpages = {8},
  year = {2020},
  month = {Feb},
  publisher = {American Physical Society},
  doi = {10.1103/PhysRevX.10.011042},
  url = {https://link.aps.org/doi/10.1103/PhysRevX.10.011042}
}

@article{aidelsburger21,
	title={Observing non-ergodicity due to kinetic constraints in tilted Fermi-Hubbard chains},
	author={Scherg, Sebastian and Kohlert, Thomas and Sala, Pablo and Pollmann, Frank and Madhusudhana, Bharath Hebbe and Bloch, Immanuel and Aidelsburger, Monika},
	journal={Nature Communications},
	volume={12},
	number={1},
	pages={1--8},
	year={2021},
	publisher={Nature Publishing Group},
    doi = {10.1038/s41467-021-24726-0}
}

@article{weitenberg22,
	title={Formation of spontaneous density-wave patterns in DC driven lattices},
	author={Zahn, HP and Singh, VP and Kosch, MN and Asteria, L and Freystatzky, L and Sengstock, K and Mathey, L and Weitenberg, C},
	journal={Physical Review X},
	volume={12},
	number={2},
	pages={021014},
	year={2022},
	publisher={APS}
}

@misc{bravyi22,
      title={Adaptive constant-depth circuits for manipulating non-abelian anyons}, 
      author={Sergey Bravyi and Isaac Kim and Alexander Kliesch and Robert Koenig},
      year={2022},
      eprint={2205.01933},
      archivePrefix={arXiv},
      primaryClass={quant-ph}
}

@article{gambetto21,
  title = {Exploiting Dynamic Quantum Circuits in a Quantum Algorithm with Superconducting Qubits},
  author = {C\'orcoles, A. D. and Takita, Maika and Inoue, Ken and Lekuch, Scott and Minev, Zlatko K. and Chow, Jerry M. and Gambetta, Jay M.},
  journal = {Phys. Rev. Lett.},
  volume = {127},
  issue = {10},
  pages = {100501},
  numpages = {6},
  year = {2021},
  month = {Aug},
  publisher = {American Physical Society},
  doi = {10.1103/PhysRevLett.127.100501},
  url = {https://link.aps.org/doi/10.1103/PhysRevLett.127.100501}
}

@article{stutz21,
  title = {Realization of Real-Time Fault-Tolerant Quantum Error Correction},
  author = {Ryan-Anderson, C. and Bohnet, J. G. and Lee, K. and Gresh, D. and Hankin, A. and Gaebler, J. P. and Francois, D. and Chernoguzov, A. and Lucchetti, D. and Brown, N. C. and Gatterman, T. M. and Halit, S. K. and Gilmore, K. and Gerber, J. A. and Neyenhuis, B. and Hayes, D. and Stutz, R. P.},
  journal = {Phys. Rev. X},
  volume = {11},
  issue = {4},
  pages = {041058},
  numpages = {29},
  year = {2021},
  month = {Dec},
  publisher = {American Physical Society},
  doi = {10.1103/PhysRevX.11.041058},
  url = {https://link.aps.org/doi/10.1103/PhysRevX.11.041058}
}

@Article{Pino21,
author={Pino, J. M.
and Dreiling, J. M.
and Figgatt, C.
and Gaebler, J. P.
and Moses, S. A.
and Allman, M. S.
and Baldwin, C. H.
and Foss-Feig, M.
and Hayes, D.
and Mayer, K.
and Ryan-Anderson, C.
and Neyenhuis, B.},
title={Demonstration of the trapped-ion quantum CCD computer architecture},
journal={Nature},
year={2021},
month={Apr},
day={01},
volume={592},
number={7853},
pages={209-213},
issn={1476-4687},
doi={10.1038/s41586-021-03318-4},
url={https://doi.org/10.1038/s41586-021-03318-4}
}

@misc{isaackim23,
      title={Experimental demonstration of the advantage of adaptive quantum circuits}, 
      author={Michael Foss-Feig and Arkin Tikku and Tsung-Cheng Lu and Karl Mayer and Mohsin Iqbal and Thomas M. Gatterman and Justin A. Gerber and Kevin Gilmore and Dan Gresh and Aaron Hankin and Nathan Hewitt and Chandler V. Horst and Mitchell Matheny and Tanner Mengle and Brian Neyenhuis and Henrik Dreyer and David Hayes and Timothy H. Hsieh and Isaac H. Kim},
      year={2023},
      eprint={2302.03029},
      archivePrefix={arXiv},
      primaryClass={quant-ph}
}

@article{aasen22,
  title = {Adiabatic paths of Hamiltonians, symmetries of topological order, and automorphism codes},
  author = {Aasen, David and Wang, Zhenghan and Hastings, Matthew B.},
  journal = {Phys. Rev. B},
  volume = {106},
  issue = {8},
  pages = {085122},
  numpages = {17},
  year = {2022},
  month = {Aug},
  publisher = {American Physical Society},
  doi = {10.1103/PhysRevB.106.085122},
  url = {https://link.aps.org/doi/10.1103/PhysRevB.106.085122}
}

@article{Hastings21,
  doi = {10.22331/q-2021-10-19-564},
  url = {https://doi.org/10.22331/q-2021-10-19-564},
  title = {Dynamically {G}enerated {L}ogical {Q}ubits},
  author = {Hastings, Matthew B. and Haah, Jeongwan},
  journal = {{Quantum}},
  issn = {2521-327X},
  publisher = {{Verein zur F{\"{o}}rderung des Open Access Publizierens in den Quantenwissenschaften}},
  volume = {5},
  pages = {564},
  month = oct,
  year = {2021}
}

@article{pace-wen,
  author = {{Pace}, Salvatore D. and {Wen}, Xiao-Gang},
  title = "{Position-dependent excitations and UV/IR mixing in the ${\mathbb{Z}}_{N}$ rank-2 toric code and its low-energy effective field theory}",
    journal = {Phys. Rev. B},
    year = 2022,
    month = Jul,
    volume = {106},
    issue = {4},
    pages = {045145},
    numpages = {20},
    doi = {10.1103/PhysRevB.106.045145}
}

@article{oh22a,
  title = {Rank-2 toric code in two dimensions},
  author = {Oh, Yun-Tak and Kim, Jintae and Moon, Eun-Gook and Han, Jung Hoon},
  journal = {Phys. Rev. B},
  volume = {105},
  issue = {4},
  pages = {045128},
  numpages = {18},
  year = {2022},
  month = {Jan},
  publisher = {American Physical Society},
  doi = {10.1103/PhysRevB.105.045128},
  url = {https://link.aps.org/doi/10.1103/PhysRevB.105.045128}
}

@article{oh22b,
  title = {Effective field theory of dipolar braiding statistics in two dimensions},
  author = {Oh, Yun-Tak and Kim, Jintae and Han, Jung Hoon},
  journal = {Phys. Rev. B},
  volume = {106},
  issue = {15},
  pages = {155150},
  numpages = {11},
  year = {2022},
  month = {Oct},
  publisher = {American Physical Society},
  doi = {10.1103/PhysRevB.106.155150},
  url = {https://link.aps.org/doi/10.1103/PhysRevB.106.155150}
}

@article{oh23,
  title = {Aspects of ${\mathbb{Z}}_{N}$ rank-2 gauge theory in $(2+1)$ dimensions: Construction schemes, holonomies, and sublattice one-form symmetries},
  author = {Oh, Yun-Tak and Pace, Salvatore D. and Han, Jung Hoon and You, Yizhi and Lee, Hyun-Yong},
  journal = {Phys. Rev. B},
  volume = {107},
  issue = {15},
  pages = {155151},
  numpages = {24},
  year = {2023},
  month = {Apr},
  publisher = {American Physical Society},
  doi = {10.1103/PhysRevB.107.155151},
  url = {https://link.aps.org/doi/10.1103/PhysRevB.107.155151}
}

@Article{iqbal24,
author={Iqbal, Mohsin
and Tantivasadakarn, Nathanan
and Verresen, Ruben
and Campbell, Sara L.
and Dreiling, Joan M.
and Figgatt, Caroline
and Gaebler, John P.
and Johansen, Jacob
and Mills, Michael
and Moses, Steven A.
and Pino, Juan M.
and Ransford, Anthony
and Rowe, Mary
and Siegfried, Peter
and Stutz, Russell P.
and Foss-Feig, Michael
and Vishwanath, Ashvin
and Dreyer, Henrik},
title={Non-Abelian topological order and anyons on a trapped-ion processor},
journal={Nature},
year={2024},
month={Feb},
day={01},
volume={626},
number={7999},
pages={505-511},
issn={1476-4687},
doi={10.1038/s41586-023-06934-4},
url={https://doi.org/10.1038/s41586-023-06934-4}
}

@article{nathanan23,
  title = {Shortest Route to Non-Abelian Topological Order on a Quantum Processor},
  author = {Tantivasadakarn, Nathanan and Verresen, Ruben and Vishwanath, Ashvin},
  journal = {Phys. Rev. Lett.},
  volume = {131},
  issue = {6},
  pages = {060405},
  numpages = {5},
  year = {2023},
  month = {Aug},
  publisher = {American Physical Society},
  doi = {10.1103/PhysRevLett.131.060405},
  url = {https://link.aps.org/doi/10.1103/PhysRevLett.131.060405}
}

@article{raussendorf01,
  title = {Persistent Entanglement in Arrays of Interacting Particles},
  author = {Briegel, Hans J. and Raussendorf, Robert},
  journal = {Phys. Rev. Lett.},
  volume = {86},
  issue = {5},
  pages = {910--913},
  numpages = {0},
  year = {2001},
  month = {Jan},
  publisher = {American Physical Society},
  doi = {10.1103/PhysRevLett.86.910},
  url = {https://link.aps.org/doi/10.1103/PhysRevLett.86.910}
}

@article{raussendorf05,
  title = {Long-range quantum entanglement in noisy cluster states},
  author = {Raussendorf, Robert and Bravyi, Sergey and Harrington, Jim},
  journal = {Phys. Rev. A},
  volume = {71},
  issue = {6},
  pages = {062313},
  numpages = {6},
  year = {2005},
  month = {Jun},
  publisher = {American Physical Society},
  doi = {10.1103/PhysRevA.71.062313},
  url = {https://link.aps.org/doi/10.1103/PhysRevA.71.062313}
}

@article{hsieh22,
  title = {Measurement as a Shortcut to Long-Range Entangled Quantum Matter},
  author = {Lu, Tsung-Cheng and Lessa, Leonardo A. and Kim, Isaac H. and Hsieh, Timothy H.},
  journal = {PRX Quantum},
  volume = {3},
  issue = {4},
  pages = {040337},
  numpages = {22},
  year = {2022},
  month = {Dec},
  publisher = {American Physical Society},
  doi = {10.1103/PRXQuantum.3.040337},
  url = {https://link.aps.org/doi/10.1103/PRXQuantum.3.040337}
}

@article{cirac21,
  title = {Quantum Circuits Assisted by Local Operations and Classical Communication: Transformations and Phases of Matter},
  author = {Piroli, Lorenzo and Styliaris, Georgios and Cirac, J. Ignacio},
  journal = {Phys. Rev. Lett.},
  volume = {127},
  issue = {22},
  pages = {220503},
  numpages = {6},
  year = {2021},
  month = {Nov},
  publisher = {American Physical Society},
  doi = {10.1103/PhysRevLett.127.220503},
  url = {https://link.aps.org/doi/10.1103/PhysRevLett.127.220503}
}

@article{verresen21a,
	title = {Long-{Range} {Entanglement} from {Measuring} {Symmetry}-{Protected} {Topological} {Phases}},
	volume = {14},
	url = {https://link.aps.org/doi/10.1103/PhysRevX.14.021040},
	doi = {10.1103/PhysRevX.14.021040},
	number = {2},
	urldate = {2024-07-31},
	journal = {Physical Review X},
	author = {Tantivasadakarn, Nathanan and Thorngren, Ryan and Vishwanath, Ashvin and Verresen, Ruben},
	month = jun,
	year = {2024},
	publisher= {American Physical Society},
	pages = {021040},
}

@misc{verresen21b,
      title={Efficiently preparing Schr\"odinger's cat, fractons and non-Abelian topological order in quantum devices}, 
      author={Ruben Verresen and Nathanan Tantivasadakarn and Ashvin Vishwanath},
      year={2022},
      eprint={2112.03061},
      archivePrefix={arXiv},
      primaryClass={quant-ph}
}

@article{seiberg23,
  title = {(2+1)-dimensional compact Lifshitz theory, tensor gauge theory, and fractons},
  author = {Gorantla, Pranay and Lam, Ho Tat and Seiberg, Nathan and Shao, Shu-Heng},
  journal = {Phys. Rev. B},
  volume = {108},
  issue = {7},
  pages = {075106},
  numpages = {24},
  year = {2023},
  month = {Aug},
  publisher = {American Physical Society},
  doi = {10.1103/PhysRevB.108.075106},
  url = {https://link.aps.org/doi/10.1103/PhysRevB.108.075106}
}

@article{nathanan20,
  title = {Searching for fracton orders via symmetry defect condensation},
  author = {Tantivasadakarn, Nathanan and Vijay, Sagar},
  journal = {Phys. Rev. B},
  volume = {101},
  issue = {16},
  pages = {165143},
  numpages = {24},
  year = {2020},
  month = {Apr},
  publisher = {American Physical Society},
  doi = {10.1103/PhysRevB.101.165143},
  url = {https://link.aps.org/doi/10.1103/PhysRevB.101.165143}
}

@article{cao24,
   title={Generating lattice non-invertible symmetries},
   volume={17},
   ISSN={2542-4653},
   url={http://dx.doi.org/10.21468/SciPostPhys.17.4.104},
   DOI={10.21468/scipostphys.17.4.104},
   number={4},
   journal={SciPost Physics},
   publisher={Stichting SciPost},
   author={Cao, Weiguang and Li, Linhao and Yamazaki, Masahito},
   year={2024},
   month=oct }

@Article{pace_gauging_2024,
	title={{Gauging modulated symmetries: Kramers-Wannier dualities and non-invertible reflections}},
	author={Salvatore D. Pace and Guilherme Delfino and Ho Tat Lam and Ömer M. Aksoy},
	journal={SciPost Phys.},
	volume={18},
	pages={021},
	year={2025},
	publisher={SciPost},
	doi={10.21468/SciPostPhys.18.1.021},
	url={https://scipost.org/10.21468/SciPostPhys.18.1.021},
}

@article{ebisu25,
    title = {Noninvertible operators in one, two, and three dimensions via gauging spatially modulated symmetry},
    volume = {111},
    url = {https://link.aps.org/doi/10.1103/PhysRevB.111.035149},
    doi = {10.1103/PhysRevB.111.035149},
    number = {3},
    urldate = {2025-03-24},
    journal = {Physical Review B},
    author = {Ebisu, Hiromi and Han, Bo},
    month = jan,
    year = {2025},
    note = {Publisher: American Physical Society},
    pages = {035149},
}

@article{bulmash_gauging_2019,
    title = {Gauging fractons: {Immobile} non-{Abelian} quasiparticles, fractals, and position-dependent degeneracies},
    volume = {100},
    shorttitle = {Gauging fractons},
    url = {https://link.aps.org/doi/10.1103/PhysRevB.100.155146},
    doi = {10.1103/PhysRevB.100.155146},
    number = {15},
    urldate = {2024-12-27},
    journal = {Physical Review B},
    author = {Bulmash, Daniel and Barkeshli, Maissam},
    month = oct,
    year = {2019},
    note = {Publisher: American Physical Society},
    pages = {155146},
}

@article{prem_gauging_2019,
    title = {Gauging permutation symmetries as a route to non-{Abelian} fractons},
    volume = {7},
    issn = {2542-4653},
    url = {https://scipost.org/10.21468/SciPostPhys.7.5.068},
    doi = {10.21468/SciPostPhys.7.5.068},
    number = {5},
    urldate = {2024-12-26},
    journal = {SciPost Physics},
    author = {Prem, Abhinav and Williamson, Dominic},
    month = nov,
    year = {2019},
    pages = {068},
}

@article{cheng_gauging_2022,
    title = {Gauging {U}(1) symmetry in (2+1)d topological phases},
    volume = {12},
    issn = {2542-4653},
    url = {https://scipost.org/10.21468/SciPostPhys.12.6.202},
    doi = {10.21468/SciPostPhys.12.6.202},
    number = {6},
    urldate = {2024-08-06},
    journal = {SciPost Physics},
    author = {Cheng, Meng and Jian, Chao-Ming},
    month = jun,
    year = {2022},
    pages = {202},
}

@article{kogut_introduction_1979,
    title = {An introduction to lattice gauge theory and spin systems},
    volume = {51},
    issn = {0034-6861},
    url = {https://link.aps.org/doi/10.1103/RevModPhys.51.659},
    doi = {10.1103/RevModPhys.51.659},
    number = {4},
    urldate = {2024-03-06},
    journal = {Reviews of Modern Physics},
    author = {Kogut, John B.},
    month = oct,
    year = {1979},
    pages = {659--713},
}

@article{mathur_lattice_2016,
    title = {Lattice gauge theories and spin models},
    volume = {94},
    issn = {2470-0010, 2470-0029},
    url = {https://link.aps.org/doi/10.1103/PhysRevD.94.085029},
    doi = {10.1103/PhysRevD.94.085029},
    number = {8},
    urldate = {2024-03-06},
    journal = {Physical Review D},
    author = {Mathur, Manu and Sreeraj, T. P.},
    month = oct,
    year = {2016},
    pages = {085029},
}

@article{shirley_foliated_2019,
    title = {Foliated fracton order from gauging subsystem symmetries},
    volume = {6},
    issn = {2542-4653},
    url = {https://scipost.org/10.21468/SciPostPhys.6.4.041},
    doi = {10.21468/SciPostPhys.6.4.041},
    number = {4},
    urldate = {2024-03-05},
    journal = {SciPost Physics},
    author = {Shirley, Wilbur and Slagle, Kevin and Chen, Xie},
    month = apr,
    year = {2019},
    pages = {041},
}

@article{kuno_interplay_2023,
    title = {Interplay between lattice gauge theory and subsystem codes},
    volume = {108},
    issn = {2469-9950, 2469-9969},
    url = {https://link.aps.org/doi/10.1103/PhysRevB.108.045150},
    doi = {10.1103/PhysRevB.108.045150},
    number = {4},
    urldate = {2024-03-05},
    journal = {Physical Review B},
    author = {Kuno, Yoshihito and Ichinose, Ikuo},
    month = jul,
    year = {2023},
    pages = {045150},
}

@article{mcgreevy_generalized_2023,
    title = {Generalized {Symmetries} in {Condensed} {Matter}},
    volume = {14},
    issn = {1947-5454, 1947-5462},
    url = {http://arxiv.org/abs/2204.03045},
    doi = {10.1146/annurev-conmatphys-040721-021029},
    number = {1},
    urldate = {2024-03-05},
    journal = {Annual Review of Condensed Matter Physics},
    author = {McGreevy, John},
    month = mar,
    year = {2023},
    note = {arXiv:2204.03045 [cond-mat, physics:hep-th]},
    pages = {57--82},
}

@misc{shao_whats_2023,
    title = {What's {Done} {Cannot} {Be} {Undone}: {TASI} {Lectures} on {Non}-{Invertible} {Symmetry}},
    shorttitle = {What's {Done} {Cannot} {Be} {Undone}},
    url = {http://arxiv.org/abs/2308.00747},
    urldate = {2024-03-12},
    publisher = {arXiv},
    author = {Shao, Shu-Heng},
    month = aug,
    year = {2023},
    note = {arXiv:2308.00747 [cond-mat, physics:hep-ph, physics:hep-th]},
}

@article{gomes_introduction_2023,
    title = {An introduction to higher-form symmetries},
    issn = {2590-1990},
    url = {https://scipost.org/10.21468/SciPostPhysLectNotes.74},
    doi = {10.21468/SciPostPhysLectNotes.74},
    urldate = {2024-06-09},
    journal = {SciPost Physics Lecture Notes},
    author = {Gomes, Pedro R. S.},
    month = sep,
    year = {2023},
    pages = {74},
}

@misc{brennan_introduction_2023,
    title = {Introduction to {Generalized} {Global} {Symmetries} in {QFT} and {Particle} {Physics}},
    url = {http://arxiv.org/abs/2306.00912},
    doi = {10.48550/arXiv.2306.00912},
    urldate = {2025-07-07},
    publisher = {arXiv},
    author = {Brennan, T. Daniel and Hong, Sungwoo},
    month = jun,
    year = {2023},
    note = {arXiv:2306.00912 [hep-ph]},
}

@article{you_higher-order_2018,
    title = {Higher-order symmetry-protected topological states for interacting bosons and fermions},
    volume = {98},
    issn = {2469-9950, 2469-9969},
    url = {https://link.aps.org/doi/10.1103/PhysRevB.98.235102},
    doi = {10.1103/PhysRevB.98.235102},
    number = {23},
    urldate = {2025-03-18},
    journal = {Physical Review B},
    author = {You, Yizhi and Devakul, Trithep and Burnell, F. J. and Neupert, Titus},
    month = dec,
    year = {2018},
    pages = {235102},
}

@article{you_higher-order_2024,
    title = {Higher-order topological phase with subsystem symmetries},
    volume = {26},
    issn = {1367-2630},
    url = {https://iopscience.iop.org/article/10.1088/1367-2630/ad78f9},
    doi = {10.1088/1367-2630/ad78f9},
    number = {9},
    urldate = {2024-09-20},
    journal = {New Journal of Physics},
    author = {You, Yizhi},
    month = sep,
    year = {2024},
    pages = {093028},
}

@article{kim23,
    title = {Unveiling {UV}/{IR} mixing via symmetry defects: {A} view from topological entanglement entropy},
    volume = {18},
    copyright = {https://creativecommons.org/licenses/by/4.0},
    issn = {2542-4653},
    shorttitle = {Unveiling {UV}/{IR} mixing via symmetry defects},
    url = {https://scipost.org/10.21468/SciPostPhys.18.3.110},
    doi = {10.21468/scipostphys.18.3.110},
    number = {3},
    urldate = {2025-07-09},
    journal = {SciPost Physics},
    author = {Kim, Jintae and Oh, Yun-Tak and Bulmash, Daniel and Han, Jung Hoon},
    month = mar,
    year = {2025},
    note = {Publisher: Stichting SciPost},
}

@misc{gorantla_string-membrane-nets_2025,
    title = {String-{Membrane}-{Nets} from {Higher}-{Form} {Gauging}: {An} {Alternate} {Route} to \$p\$-{String} {Condensation}},
    shorttitle = {String-{Membrane}-{Nets} from {Higher}-{Form} {Gauging}},
    url = {http://arxiv.org/abs/2505.13604},
    doi = {10.48550/arXiv.2505.13604},
    urldate = {2025-05-29},
    publisher = {arXiv},
    author = {Gorantla, Pranay and Prem, Abhinav and Tantivasadakarn, Nathanan and Williamson, Dominic J.},
    month = may,
    year = {2025},
    note = {arXiv:2505.13604 [cond-mat]},
}

@misc{li_non-invertible_2024,
    title = {Non-invertible {SPT}, gauging and symmetry fractionalization},
    url = {http://arxiv.org/abs/2405.15951},
    doi = {10.48550/arXiv.2405.15951},
    urldate = {2025-04-14},
    publisher = {arXiv},
    author = {Li, Yabo and Litvinov, Mikhail},
    month = may,
    year = {2024},
    note = {arXiv:2405.15951 [cond-mat]},
}

@misc{seifnashri_gauging_2025,
    title = {Gauging non-invertible symmetries on the lattice},
    url = {http://arxiv.org/abs/2503.02925},
    doi = {10.48550/arXiv.2503.02925},
    urldate = {2025-03-07},
    publisher = {arXiv},
    author = {Seifnashri, Sahand and Shao, Shu-Heng and Yang, Xinping},
    month = mar,
    year = {2025},
    note = {arXiv:2503.02925 [cond-mat]},
}

@misc{lyons_protocols_2024,
    title = {Protocols for {Creating} {Anyons} and {Defects} via {Gauging}},
    url = {http://arxiv.org/abs/2411.04181},
    doi = {10.48550/arXiv.2411.04181},
    urldate = {2024-11-29},
    publisher = {arXiv},
    author = {Lyons, Anasuya and Lo, Chiu Fan Bowen and Tantivasadakarn, Nathanan and Vishwanath, Ashvin and Verresen, Ruben},
    month = nov,
    year = {2024},
    note = {arXiv:2411.04181},
}

@article{lam24,
  title = {Classification of dipolar symmetry-protected topological phases: Matrix product states, stabilizer Hamiltonians, and finite tensor gauge theories},
  author = {Lam, Ho Tat},
  journal = {Phys. Rev. B},
  volume = {109},
  issue = {11},
  pages = {115142},
  numpages = {21},
  year = {2024},
  month = {Mar},
  publisher = {American Physical Society},
  doi = {10.1103/PhysRevB.109.115142},
  url = {https://link.aps.org/doi/10.1103/PhysRevB.109.115142}
}

@article{you24,
  title = {Intrinsic symmetry-protected topological mixed state from modulated symmetries and hierarchical structure of boundary anomaly},
  author = {You, Yizhi and Oshikawa, Masaki},
  journal = {Phys. Rev. B},
  volume = {110},
  issue = {16},
  pages = {165160},
  numpages = {15},
  year = {2024},
  month = {Oct},
  publisher = {American Physical Society},
  doi = {10.1103/PhysRevB.110.165160},
  url = {https://link.aps.org/doi/10.1103/PhysRevB.110.165160}
}

@article{bhardwaj24,
title = {Lectures on generalized symmetries},
journal = {Physics Reports},
volume = {1051},
pages = {1-87},
year = {2024},
issn = {0370-1573},
doi = {https://doi.org/10.1016/j.physrep.2023.11.002},
url = {https://www.sciencedirect.com/science/article/pii/S0370157323003861},
author = {Lakshya Bhardwaj and Lea E. Bottini and Ludovic Fraser-Taliente and Liam Gladden and Dewi S.W. Gould and Arthur Platschorre and Hannah Tillim}
}

@article{kim_model_2021,
    title = {Model for fractons, fluxons, and free vertex excitations},
    volume = {104},
    issn = {2469-9950, 2469-9969},
    url = {https://link.aps.org/doi/10.1103/PhysRevB.104.115128},
    doi = {10.1103/PhysRevB.104.115128},
    number = {11},
    urldate = {2024-03-11},
    journal = {Physical Review B},
    author = {Kim, Jintae and Han, Jung Hoon},
    month = sep,
    year = {2021},
    pages = {115128},
}

@article{seifnashri_cluster_2024,
    title = {Cluster {State} as a {Noninvertible} {Symmetry}-{Protected} {Topological} {Phase}},
    volume = {133},
    url = {https://link.aps.org/doi/10.1103/PhysRevLett.133.116601},
    doi = {10.1103/PhysRevLett.133.116601},
    number = {11},
    urldate = {2024-10-17},
    journal = {Physical Review Letters},
    author = {Seifnashri, Sahand and Shao, Shu-Heng},
    month = sep,
    year = {2024},
    note = {Publisher: American Physical Society},
    pages = {116601},
}

@misc{lee_decoding_2022,
    title = {Decoding {Measurement}-{Prepared} {Quantum} {Phases} and {Transitions}: from {Ising} model to gauge theory, and beyond},
    shorttitle = {Decoding {Measurement}-{Prepared} {Quantum} {Phases} and {Transitions}},
    url = {http://arxiv.org/abs/2208.11699},
    urldate = {2024-03-05},
    publisher = {arXiv},
    author = {Lee, Jong Yeon and Ji, Wenjie and Bi, Zhen and Fisher, Matthew P. A.},
    month = sep,
    year = {2022},
    note = {arXiv:2208.11699 [cond-mat, physics:quant-ph]},
}

@article{xu_entanglement_2025,
    title = {Entanglement {Properties} of {Gauge} {Theories} from {Higher}-{Form} {Symmetries}},
    volume = {15},
    issn = {2160-3308},
    url = {https://link.aps.org/doi/10.1103/PhysRevX.15.011001},
    doi = {10.1103/PhysRevX.15.011001},
    number = {1},
    urldate = {2025-07-08},
    journal = {Physical Review X},
    author = {Xu, Wen-Tao and Rakovszky, Tibor and Knap, Michael and Pollmann, Frank},
    month = jan,
    year = {2025},
    pages = {011001},
}

@article{choi_non-invertible_nodate,
    title={{Non-invertible and higher-form symmetries in 2+1d lattice gauge theories}},
	author={Yichul Choi and Yaman Sanghavi and Shu-Heng Shao and Yunqin Zheng},
	journal={SciPost Phys.},
	volume={18},
	pages={008},
	year={2025},
	publisher={SciPost},
	doi={10.21468/SciPostPhys.18.1.008},
	url={https://scipost.org/10.21468/SciPostPhys.18.1.008},
}

@article{li_symmetry-enriched_2023,
    title = {Symmetry-enriched topological order from partially gauging symmetry-protected topologically ordered states assisted by measurements},
    volume = {108},
    url = {https://link.aps.org/doi/10.1103/PhysRevB.108.115144},
    doi = {10.1103/PhysRevB.108.115144},
    number = {11},
    urldate = {2024-03-06},
    journal = {Physical Review B},
    author = {Li, Yabo and Sukeno, Hiroki and Mana, Aswin Parayil and Nautrup, Hendrik Poulsen and Wei, Tzu-Chieh},
    month = sep,
    year = {2023},
    note = {Publisher: American Physical Society},
    pages = {115144},
}

@misc{yoshitome_generalized_2025,
    title = {Generalized {Modulated} {Symmetries} in \${\textbackslash}mathbb\{{Z}\}\_2\$ {Topological} {Ordered} {Phases}},
    url = {http://arxiv.org/abs/2506.10819},
    doi = {10.48550/arXiv.2506.10819},
    urldate = {2025-06-24},
    publisher = {arXiv},
    author = {Yoshitome, Gustavo M. and Casasola, Heitor and Corso, Rodrigo and Gomes, Pedro R. S.},
    month = jun,
    year = {2025},
    note = {arXiv:2506.10819 [cond-mat]},
}

@misc{gorantla_tensor_2024,
    title = {Tensor networks for non-invertible symmetries in 3+1d and beyond},
    url = {http://arxiv.org/abs/2406.12978},
    doi = {10.48550/arXiv.2406.12978},
    urldate = {2024-12-09},
    publisher = {arXiv},
    author = {Gorantla, Pranay and Shao, Shu-Heng and Tantivasadakarn, Nathanan},
    month = jun,
    year = {2024},
    note = {arXiv:2406.12978},
}

@article{kim_hybrid_2022,
    title = {Hybrid rank-1 and rank-2 {U}(1) lattice gauge theory, the {F3} model, and its effective field theory},
    volume = {106},
    issn = {2469-9950, 2469-9969},
    url = {https://link.aps.org/doi/10.1103/PhysRevB.106.155154},
    doi = {10.1103/PhysRevB.106.155154},
    number = {15},
    urldate = {2024-03-11},
    journal = {Physical Review B},
    author = {Kim, Jintae and Oh, Yun-Tak and Han, Jung Hoon},
    month = oct,
    year = {2022},
    pages = {155154},
}

@article{ebisu_foliated_2024,
    title = {Foliated field theories and multipole symmetries},
    volume = {109},
    url = {https://link.aps.org/doi/10.1103/PhysRevB.109.165112},
    doi = {10.1103/PhysRevB.109.165112},
    number = {16},
    urldate = {2025-06-28},
    journal = {Physical Review B},
    author = {Ebisu, Hiromi and Honda, Masazumi and Nakanishi, Taiichi},
    month = apr,
    year = {2024},
    note = {Publisher: American Physical Society},
    pages = {165112},
}

@article{seiberg_exotic_2020,
    title = {Exotic \${U}(1)\$ symmetries, duality, and fractons in 3+1-dimensional quantum field theory},
    volume = {9},
    issn = {2542-4653},
    url = {https://scipost.org/10.21468/SciPostPhys.9.4.046},
    doi = {10.21468/SciPostPhys.9.4.046},
    number = {4},
    urldate = {2024-03-06},
    journal = {SciPost Physics},
    author = {Seiberg, Nathan and Shao, Shu-Heng},
    month = oct,
    year = {2020},
    pages = {046},
}

@article{seiberg_exotic_2021,
    title = {Exotic symmetries, duality, and fractons in 2+1-dimensional quantum field theory},
    volume = {10},
    issn = {2542-4653},
    url = {https://scipost.org/10.21468/SciPostPhys.10.2.027},
    doi = {10.21468/SciPostPhys.10.2.027},
    number = {2},
    urldate = {2024-03-05},
    journal = {SciPost Physics},
    author = {Seiberg, Nathan and Shao, Shu-Heng},
    month = feb,
    year = {2021},
    pages = {027},
}

@misc{bulmash_defect_2025,
    title = {Defect {Networks} for {Topological} {Phases} {Protected} {By} {Modulated} {Symmetries}},
    url = {http://arxiv.org/abs/2508.06604},
    doi = {10.48550/arXiv.2508.06604},
    urldate = {2025-08-13},
    publisher = {arXiv},
    author = {Bulmash, Daniel},
    month = aug,
    year = {2025},
    note = {arXiv:2508.06604 [cond-mat]},
}

@article{ren_efficient_2025,
    title = {Efficient {Preparation} of {Solvable} {Anyons} with {Adaptive} {Quantum} {Circuits}},
    volume = {15},
    issn = {2160-3308},
    url = {https://link.aps.org/doi/10.1103/b9hf-gx4f},
    doi = {10.1103/b9hf-gx4f},
    number = {3},
    urldate = {2025-09-03},
    journal = {Physical Review X},
    author = {Ren, Yuanjie and Tantivasadakarn, Nathanan and Williamson, Dominic J.},
    month = aug,
    year = {2025},
    pages = {031060},
}

@misc{kim25,
      title={Noninvertible symmetry and topological holography for modulated SPT in one dimension}, 
      author={Jintae Kim and Yizhi You and Jung Hoon Han},
      year={2025},
      eprint={2507.02324},
      archivePrefix={arXiv},
      primaryClass={cond-mat.str-el},
      url={https://arxiv.org/abs/2507.02324}, 
}

@dataset{kim25zenodo,
  author       = {Kim, Jintae},
  title        = {Source Codes and Numerical Calculation Data for
                   "From Paramagnet to Dipolar Topological Order via
                   Duality and Dipolar SPT"
                  },
  month        = apr,
  year         = 2025,
  publisher    = {Zenodo},
  doi          = {10.5281/zenodo.15165917},
  url          = {https://doi.org/10.5281/zenodo.15165917},
}

\end{document}